%% file: Manuscript.tex
\documentclass[twocolumn]{aastex631}

\input{Commands}

\input{Packages}

\begin{document}

\newcommand{\cntext}[1]{\begin{CJK}{UTF8}{bsmi}#1\end{CJK}}
\newcommand{\cnstext}[1]{\begin{CJK}{UTF8}{gbsn}#1\end{CJK}}

\title{Discovery and Extensive Follow-Up of SN~2024ggi, a nearby type IIP supernova in NGC~3621}

\correspondingauthor{T.-W.~Chen}
\email{twchen@astro.ncu.edu.tw}

\correspondingauthor{S.~Yang}
\email{sheng.yang@hnas.ac.cn}

\author[0000-0002-1066-6098]{Ting-Wan~Chen \cntext{(陳婷琬)}} 
\affil{Graduate Institute of Astronomy, National Central University, 300 Jhongda Road, 32001 Jhongli, Taiwan}

\author[0000-0002-2898-6532]{Sheng~Yang \cnstext{(杨圣)}} 
\affil{Henan Academy of Sciences, Zhengzhou 450046, Henan, China}

\author[0000-0003-4524-6883]{Shubham~Srivastav}
\affil{Astrophysics sub-Department, Department of Physics, University of Oxford, Keble Road, Oxford, OX1 3RH, UK}

\author[0000-0003-1169-1954]{Takashi~J.~Moriya}
\affiliation{National Astronomical Observatory of Japan, National Institutes of Natural Sciences, 2-21-1 Osawa, Mitaka, Tokyo 181-8588, Japan}
\affiliation{Graduate Institute for Advanced Studies, SOKENDAI, 2-21-1 Osawa, Mitaka, Tokyo 181-8588, Japan}
\affiliation{School of Physics and Astronomy, Monash University, Clayton, VIC 3800, Australia}

\author[0000-0002-8229-1731]{Stephen~J.~Smartt}
\affil{Astrophysics sub-Department, Department of Physics, University of Oxford, Keble Road, Oxford, OX1 3RH, UK}
\affil{Astrophysics Research Centre, School of Mathematics and Physics, Queen's University Belfast, BT7 1NN, UK}

\author[0000-0002-3825-0553]{Sofia~Rest}
\affil{Department of Computer Science, The Johns Hopkins University, Baltimore, MD 21218, USA}

\author{Armin~Rest}
\affil{Space Telescope Science Institute, 3700 San Martin Drive, Baltimore, MD 21218, USA}
\affil{Department of Physics and Astronomy, Johns Hopkins University, Baltimore, MD 21218, USA}

\author[0000-0001-7737-6784]{Hsing~Wen~Lin \cntext{(林省文)}} 
\affil{Department of Physics, University of Michigan, 450 Church Street, Ann Arbor, MI 48109-1107, USA}

\author[0000-0003-2736-5977]{Hao-Yu~Miao \cntext{(繆皓宇)}} 
\affil{Graduate Institute of Astronomy, National Central University, 300 Jhongda Road, 32001 Jhongli, Taiwan}

\author[0000-0001-9686-5874]{Yu-Chi~Cheng \cntext{(鄭宇棋)}} 
\affiliation{Department of Physics, National Taiwan Normal University, No. 88, Sect. 4, Tingzhou Rd., Wenshan Dist., Taipei City, 116325 Taiwan} 
\affiliation{Center of Astronomy and Gravitation, National Taiwan Normal University, No. 88, Sect. 4, Tingzhou Rd., Wenshan Dist., Taipei City, 116325 Taiwan}

\author[0000-0002-9928-0369]{Amar~Aryan} 
\affil{Graduate Institute of Astronomy, National Central University, 300 Jhongda Road, 32001 Jhongli, Taiwan}

\author{Chia-Yu~Cheng \cntext{(鄭家羽)}} 
\affil{Graduate Institute of Astronomy, National Central University, 300 Jhongda Road, 32001 Jhongli, Taiwan}

\author[0000-0003-2191-1674]{Morgan~Fraser}
\affil{School of Physics, O'Brien Centre for Science North, University College Dublin, Belfield, Dublin 4, Ireland}

\author[0000-0002-9679-5279]{Li-Ching~Huang \cntext{(黃立晴)}}  
\affiliation{Department of Physics, National Taiwan Normal University, No. 88, Sect. 4, Tingzhou Rd., Wenshan Dist., Taipei City, 116325 Taiwan} 
\affiliation{Center of Astronomy and Gravitation, National Taiwan Normal University, No. 88, Sect. 4, Tingzhou Rd., Wenshan Dist., Taipei City, 116325 Taiwan}

\author{Meng-Han~Lee \cntext{(李孟翰)}} 
\affil{Graduate Institute of Astronomy, National Central University, 300 Jhongda Road, 32001 Jhongli, Taiwan}

\author{Cheng-Han~Lai \cntext{(賴政翰)}} 
\affil{Graduate Institute of Astronomy, National Central University, 300 Jhongda Road, 32001 Jhongli, Taiwan}

\author[0009-0005-2378-2601]{Yu-Hsuan~Liu \cntext{(劉宇軒)}} 
\affil{Graduate Institute of Astronomy, National Central University, 300 Jhongda Road, 32001 Jhongli, Taiwan}
\affil{Institute of Astronomy and Astrophysics, Academia Sinica, No.1, Sec. 4, Roosevelt Rd., Taipei 106216, Taiwan}

\author{Aiswarya~Sankar.K} 
\affil{Graduate Institute of Astronomy, National Central University, 300 Jhongda Road, 32001 Jhongli, Taiwan}

\author[0000-0001-9535-3199]{Ken~W.~Smith} 
\affil{Astrophysics Research Centre, School of Mathematics and Physics, Queen's University Belfast, BT7 1NN, UK}

\author[0000-0002-0504-4323]{Heloise~F.~Stevance}
\affil{Astrophysics sub-Department, Department of Physics, University of Oxford, Keble Road, Oxford, OX1 3RH, UK}
\affil{Astrophysics Research Centre, School of Mathematics and Physics, Queen's University Belfast, BT7 1NN, UK}

\author{Ze-Ning~Wang \cnstext{(王泽宁)}} 
\affil{Henan Academy of Sciences, Zhengzhou 450046, Henan, China}
\affil{School of Physics, Henan Normal University, Xinxiang 453007, Henan, China}

\author[0000-0003-0227-3451]{Joseph~P.~Anderson} 
\affil{European Southern Observatory, Alonso de C{\'o}rdova 3107, Casilla 19, Santiago, Chile}
\affil{Millennium Institute of Astrophysics MAS, Nuncio Monse{\~n}or Sotero Sanz 100, Off. 104, Providencia, Santiago, Chile}

\author[0000-0002-4269-7999]{Charlotte~R.~Angus} 
\affil{Astrophysics Research Centre, School of Mathematics and Physics, Queens University Belfast, Belfast BT7 1NN, UK}

\author{Thomas~de~Boer} 
\affil{Institute for Astronomy, University of Hawai'i, 2680 Woodlawn Drive, Honolulu, HI 96822, USA}

\author{Kenneth~Chambers} 
\affil{Institute for Astronomy, University of Hawai'i, 2680 Woodlawn Drive, Honolulu, HI 96822, USA}

\author[0000-0002-7022-4742]{Hao-Yuan~Duan \cntext{(段皓元)}}  
\affil{Taipei Astronomical Museum, Taipei, Taiwan}

\author[0000-0002-9986-3898]{Nicolas~Erasmus} 
\affil{South African Astronomical Observatory, 1 Observatory Road, Cape Town, 7925, South Africa}

\author[0000-0003-1015-5367]{Hua~Gao} 
\affil{Institute for Astronomy, University of Hawai'i, 2680 Woodlawn Drive, Honolulu, HI 96822, USA}

\author{Joanna~Herman} 
\affil{Institute for Astronomy, University of Hawai'i, 2680 Woodlawn Drive, Honolulu, HI 96822, USA}

\author{Wei-Jie~Hou \cntext{(侯偉傑)}} 
\affil{Graduate Institute of Astronomy, National Central University, 300 Jhongda Road, 32001 Jhongli, Taiwan}

\author{Hsiang-Yao~Hsiao \cntext{(蕭翔耀)}} 
\affil{Graduate Institute of Astronomy, National Central University, 300 Jhongda Road, 32001 Jhongli, Taiwan}

\author[0000-0003-1059-9603]{Mark~E.~Huber} 
\affil{Institute for Astronomy, University of Hawai'i, 2680 Woodlawn Drive, Honolulu, HI 96822, USA}

\author[0000-0002-7272-5129]{Chien-Cheng~Lin \cntext{(林建爭)}} 
\affil{Institute for Astronomy, University of Hawai'i, 2680 Woodlawn Drive, Honolulu, HI 96822, USA}

\author{Hung-Chin~Lin \cntext{(林宏欽)}} 
\affil{Graduate Institute of Astronomy, National Central University, 300 Jhongda Road, 32001 Jhongli, Taiwan}

\author{Eugene~A.~Magnier} 
\affil{Institute for Astronomy, University of Hawai'i, 2680 Woodlawn Drive, Honolulu, HI 96822, USA}

\author[0000-0001-7735-928X]{Ka~Kit~Man \cntext{(文家傑)}}   
\affil{School of Business, Hong Kong Baptist University}

\author[0000-0001-8385-3727]{Thomas~Moore} 
\affil{Astrophysics Research Centre, School of Mathematics and Physics, Queens University Belfast, Belfast BT7 1NN, UK}
\affil{European Southern Observatory, Alonso de C\'{o}rdova 3107, Casilla 19, Santiago, Chile}

\author[0000-0001-8771-7554]{Chow-Choong~Ngeow \cntext{(饒兆聰)}} 
\affil{Graduate Institute of Astronomy, National Central University, 300 Jhongda Road, 32001 Jhongli, Taiwan}

\author[0000-0002-2555-3192]{Matt~Nicholl} 
\affil{Astrophysics Research Centre, School of Mathematics and Physics, Queens University Belfast, Belfast BT7 1NN, UK}

\author[0000-0003-1295-8235]{Po-Sheng~Ou \cntext{(歐柏昇)}}  
\affil{Department of Physics, National Taiwan University, No.1, Sec. 4, Roosevelt Rd., Taipei 106216, Taiwan}
\affil{Institute of Astronomy and Astrophysics, Academia Sinica, No.1, Sec. 4, Roosevelt Rd., Taipei 106216, Taiwan}

\author{Giuliano~Pignata} 
\affil{Instituto de Alta Investigaci{\'o}n, Universidad de Tarapac{\'a}, Casilla 7D, Arica, Chile}
\affil{Millennium Institute of Astrophysics MAS, Nuncio Monse{\~n}or Sotero Sanz 100, Off. 104, Providencia, Santiago, Chile}

\author{Yu-Chien~Shiau \cntext{(蕭聿謙)}}  
\affil{Taipei Amateur Astronomers Association}

\author[0000-0002-1154-8317]{Julian~Silvester~Sommer} 
\affil{Universit{\"a}ts-Sternwarte, Fakult{\"a}t f{\"u}r Physik, Ludwig-Maximilians Universit{\"a}t, Scheinerstr. 1, 81679 M{\"u}nchen, Germany}

\author[0000-0003-2858-9657]{John~L.~Tonry} 
\affil{Institute for Astronomy, University of Hawai'i, 2680 Woodlawn Drive, Honolulu, HI 96822, USA}

\author[0000-0002-7334-2357]{Xiao-Feng~Wang \cnstext{(王晓锋)}} 
\affiliation{Physics Department, Tsinghua University, Beijing, 100084, China}
\affiliation{Purple Mountain Observatory, Chinese Academy of Science, Nanjing, 210023, China}

\author[0000-0002-1229-2499]{David~R.~Young} 
\affil{Astrophysics Research Centre, School of Mathematics and Physics, Queens University Belfast, Belfast BT7 1NN, UK}

\author{You-Ting~Yeh \cntext{(葉祐廷)}}  
\affil{CheCheng Elementary School Observatory}

\author[0000-0002-8296-2590]{Jujia~Zhang \cnstext{(张居甲)}} 
\affiliation{Yunnan Observatories (YNAO), Chinese Academy of Sciences (CAS), Kunming, 650216, China}
\affiliation{International Centre of Supernovae, Yunnan Key Laboratory, Kunming 650216, China}

\begin{abstract}
We present the discovery and early observations of the nearby Type II supernova (SN) 2024ggi in NGC~3621 at $6.64\pm0.3$\,Mpc. The SN was caught $5.8^{+1.9}_{-2.9}$ hours after its explosion by the ATLAS survey. Early-phase, high-cadence, and multi-band photometric follow-up was performed by the Kinder (Kilonova Finder) project, collecting over 1000 photometric data points within a week. The combined $o$- and $r$-band light curves show a rapid rise of 3.3 magnitudes in 13.7 hours, much faster than SN~2023ixf (another recent, nearby, and well-observed SN~II). Between 13.8 and 18.8 hours after explosion SN~2024ggi became bluer, with $u-g$ colour dropping from 0.53 to 0.15\,mag. The rapid blueward evolution indicates a wind shock breakout (SBO) scenario. 
No hour-long brightening expected for the SBO from a bare stellar surface was detected during our observations.  
The classification spectrum, taken 17 hours after the SN explosion, shows flash features of high-ionization species such as Balmer lines, He~\I, C~\III, and N~\III. Detailed light curve modeling reveals critical insights into the properties of the circumstellar material (CSM). 
Our favoured model has an explosion energy of $2\times 10^{51}~\mathrm{erg}$, a mass-loss rate of $10^{-3}~\mathrm{M_\odot~yr^{-1}}$ (with an assumed $10~\mathrm{km~s^{-1}}$ wind), and a confined CSM radius of $6\times 10^{14}~\mathrm{cm}$. The corresponding CSM mass is $0.4~\mathrm{M_\odot}$. 
Comparisons with SN~2023ixf highlight that SN~2024ggi has a smaller CSM density, resulting in a faster rise and fainter UV flux. The extensive dataset and the involvement of citizen astronomers underscore that a collaborative network is essential for SBO searches, leading to more precise and comprehensive SN characterizations.
\end{abstract}

\keywords{(stars:) supernovae: general, (stars:) supernovae: individual (SN~2024ggi)}

\section{Introduction} \label{sec:intro}
Core-collapse supernovae (SNe) mark the end stage of massive ($>8~\mathrm{M_\odot}$) star evolution. Those that have hydrogen present in their spectra are classified as Type II SNe. Within this category, some display a plateau in their light curves (Type IIP), which is thought to result from the hydrogen recombination in the expanding ejecta.
Capturing such events as early as possible is crucial for detecting the phenomenon known as shock breakout (SBO), where the SN shock wave emerges from the stellar surface or from a dense circumstellar material (CSM) surrounding the progenitor. 
If the CSM is not dense, this SBO event can be marked by a brief but intense burst of high-energy radiation, providing direct insights into the outermost layers of the progenitor star \citep[e.g.,][]{2017hsn..book..967W}.
If the CSM is dense and optically thick enough, SBO occurs within the CSM. In such a case, we do not expect to observe the brief but intense burst because of photon diffusion in the dense CSM, but the early optical light curves evolve more quickly \citep[e.g.,][]{2011MNRAS.415..199M,2018MNRAS.476.2840M,2017ApJ...838...28M,2018ApJ...858...15M}. Such a quick rise is often observed in SNe~II \citep[e.g.,][]{2015MNRAS.451.2212G,2018NatAs...2..808F}. 
\citet{2022ApJ...933..164G} suggest a reason for the absence of a detected SBO signature can be attributed to the realistic 3D structure of a RSG, which includes large convective bubbles at the photosphere (as observed in Betelgeuse). This structure spreads the SBO over a longer duration, resulting in a fainter signal.

If caught early, SNe II spectra can show narrow emission lines of high-ionization species \citep[e.g.][]{1985ApJ...289...52N,2014Natur.509..471G}. These ``flash'' features disappear within hours to days, and result from the interaction of the SN shock wave with CSM surrounding the progenitor star. These flash spectra provide a detailed view of the immediate environment of the progenitor star and the recent mass-loss history in the narrow window between core collapse and the ejecta sweeping up the immediate surroundings. Sample studies suggest that more than 40\% of SNe II discovered within two days of first light show flash features from interaction with dense CSM \citep{2021ApJ...912...46B,2023ApJ...952..119B}.
SN~2023ixf is a notable recent example of a 
flash SN~IIP, discovered early by a citizen astronomer \citep{2023TNSTR1158....1I} and monitored extensively 
\citep[e.g.][]{2023SciBu..68.2548Z,2023ApJ...956L...5B}.

SN~2024ggi is one of the nearest SNe of the present decade \citep[full details see Sec.\,\ref{sec:obs}]{2024TNSTR1020....1T,2024TNSAN.100....1S}, following SN~2014J in M82 \citep[Type Ia;][]{2014CBET.3792....1F} and SN~2023ixf in M101. 
Its host galaxy NGC~3621 is a well-known spiral galaxy located only 6.64\,Mpc away (more details see Sec.\,\ref{sec:analysis_host}). 
SN~2024ggi was discovered very early (more details see Sec.\,\ref{sec:analysis_explosion_sbo}) by ATLAS and has been exceptionally well-monitored with multi-wavelength observations: $\gamma$-ray \citep{2024ATel16601....1M}, X-ray \citep{2024ATel16588....1Z}, optical \citep{2024TNSAN.101....1K,2024TNSAN.102....1C,2024TNSAN.108....1K,2024TNSAN.109....1R}, radio \citep{2024ATel16616....1R}, and limits at centimeter-wavelength \citep{2024ATel16612....1C}.
\citet{2024TNSAN.103....1H} and \citet{2024TNSAN.104....1Z} obtained spectra for SN~2024ggi, classifying it as a young Type II SN with flash ionization features (we present one classification spectrum in Sec.\,\ref{sec:spec}). 
Further well-analysed datasets of early spectroscopy within one-two days after the SN discovery have been presented in \citet{2024arXiv240419006J,2024arXiv240502274P,2024arXiv240518490S}. 
These spectra displayed strong and narrow features of high-ionization species including He~\I, He~\II, N~\III, C~\III, N~\IV\ and C~\IV. 
Later on, a rise in ionization was also observed as indicated by the presence of He~\II, C~\IV, N~\IV/\V\ and O~\V\ features.  
Several groups searched for the progenitor star using the archival imaging from the Hubble Space Telescope (HST), Dark Energy Camera Legacy Survey (DECaLS) and XMM-Newton  \citep{2024TNSAN.100....1S,2024TNSAN.105....1Y,2024TNSAN.107....1P,2024ATel16595....1K,2024arXiv240507964C}. 
Utilising the pre-explosion images from HST and {\textit Spitzer} Space Telescope, \citet[][]{xiang2024red} found that SN~2024ggi probably resulted from the explosion of a solar metallicity massive star having an initial mass of 13$^{+1}_{-1}$\,M$_{\odot}$. 

In this Letter, we report the discovery of SN~2024ggi, and focus on its early photometric and spectroscopic observations in Section\,\ref{sec:obs}, including images contributed by citizen scientists. 
In Section\,\ref{sec:analysis}, we highlight the importance of early detection and high-cadence monitoring for uncovering details of SN~2024ggi, such as searching precursors and SBO signals, constructing a bolometric light curve, temperature and radius evolution. 
In Section\,\ref{sec:discussions}, we model the light curve to estimate CSM properties, discuss our results, and compare the properties of SN~2024ggi to several classical Type II SNe, particularly SN2023ixf. 
Finally we conclude our findings in Section\,\ref{sec:conclusion}. 
Throughout this paper, all magnitudes are reported in the AB system.

\section{Observations} 
\label{sec:obs}

\subsection{ATLAS discovery of SN~2024ggi}
The Asteroid Terrestrial-impact Last Alert System \citep[ATLAS;][]{2018PASP..130f4505T} comprises a network of four 0.5-meter telescopes located in Hawaii, Chile, and South Africa, facilitating wide-field all-sky observations. 
These telescopes continuously scan the entire visible sky, completing approximately four scans within a 24-hour period when all units are operational. Following data acquisition, automated image processing occurs \citep{2018PASP..130f4505T} incorporating photometric and astrometric calibration procedures using the reference catalogue RefCat2 \citep{2018ApJ...867..105T}. Subsequently, a reference image is subtracted to find transient events. 
The significant sources detected on the difference images are filtered through a transient
discovery pipeline 
\citep[the ATLAS Transient Server;][]{Smith2020}. 
This streamlined process enables rapid identification of extragalactic transients, and 
all data can be accessed\footnote{https://fallingstar.com} through our forced photometry server \citep{2021TNSAN...7....1S}. 

\begin{figure*}
    \centering
    \subfigure{\includegraphics[width=0.4\textwidth]{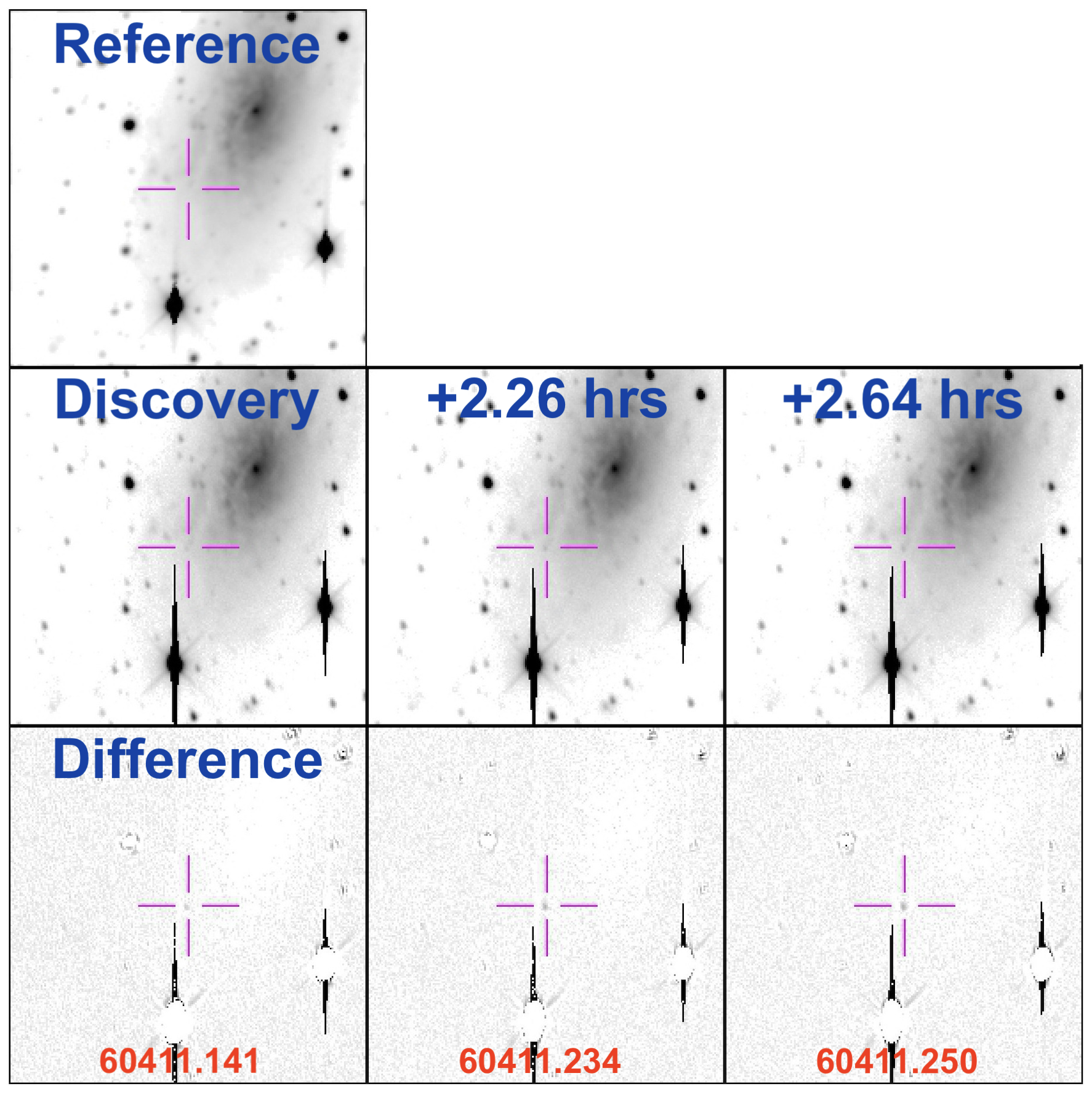}}
    \subfigure{\includegraphics[width=0.59\textwidth]{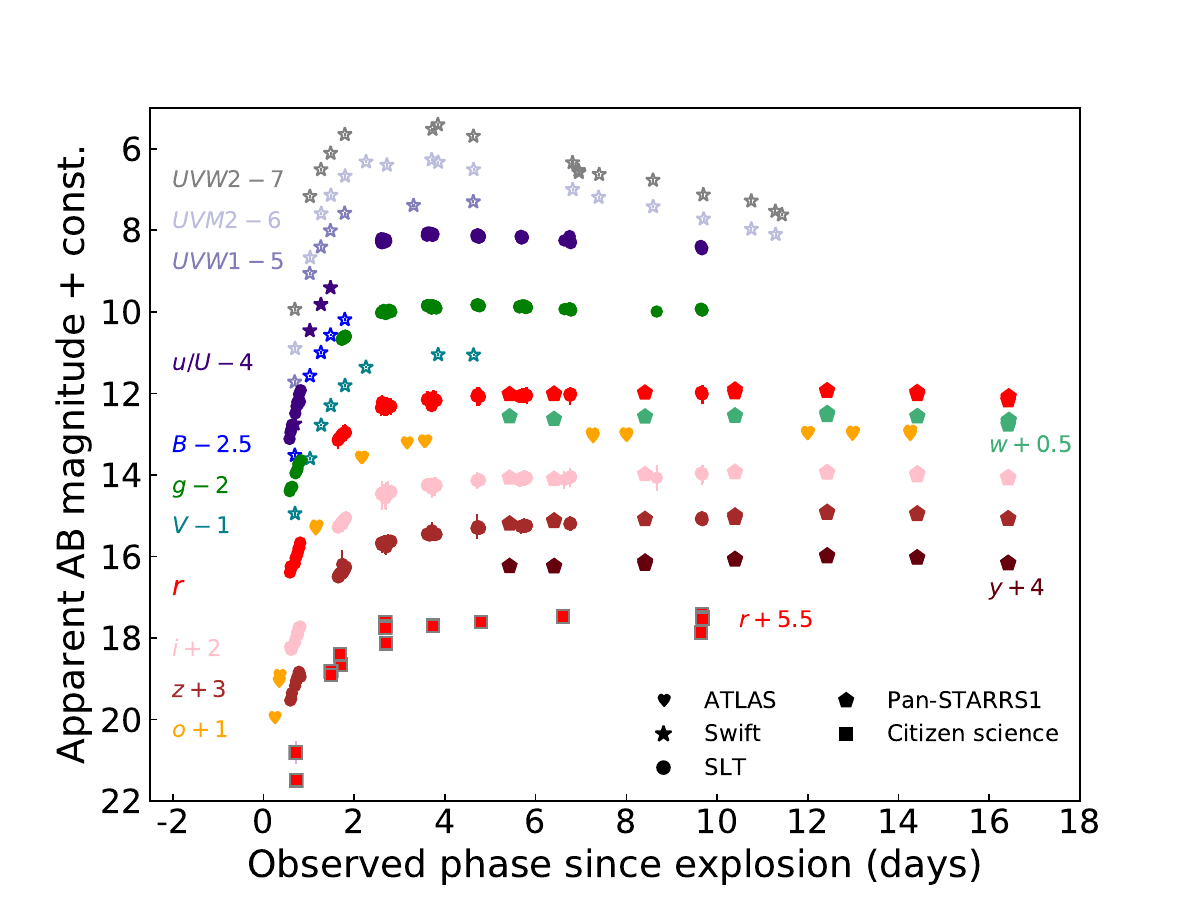}}
    \subfigure{\includegraphics[width=0.4\textwidth]{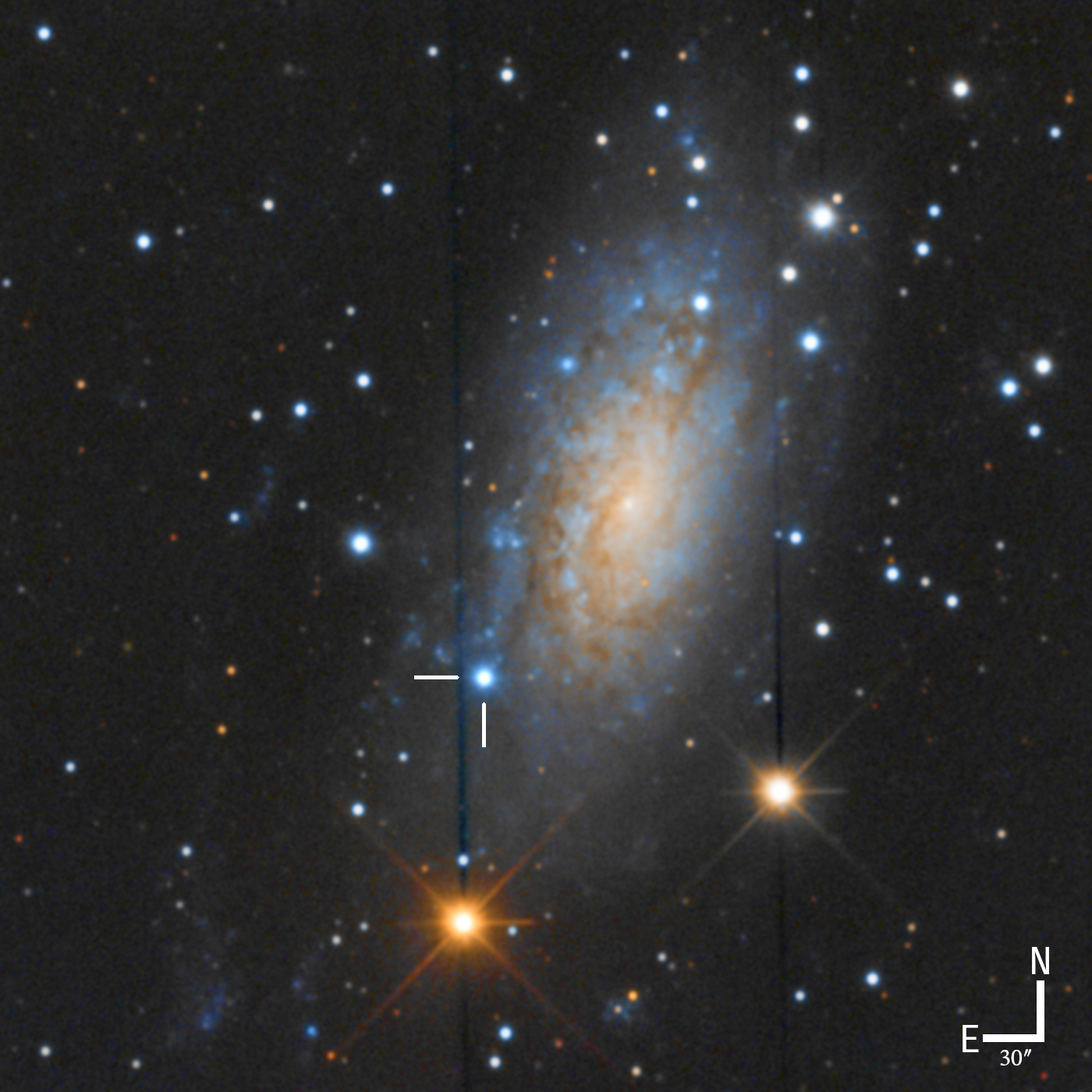}}
    \subfigure{\includegraphics[width=0.59\textwidth]{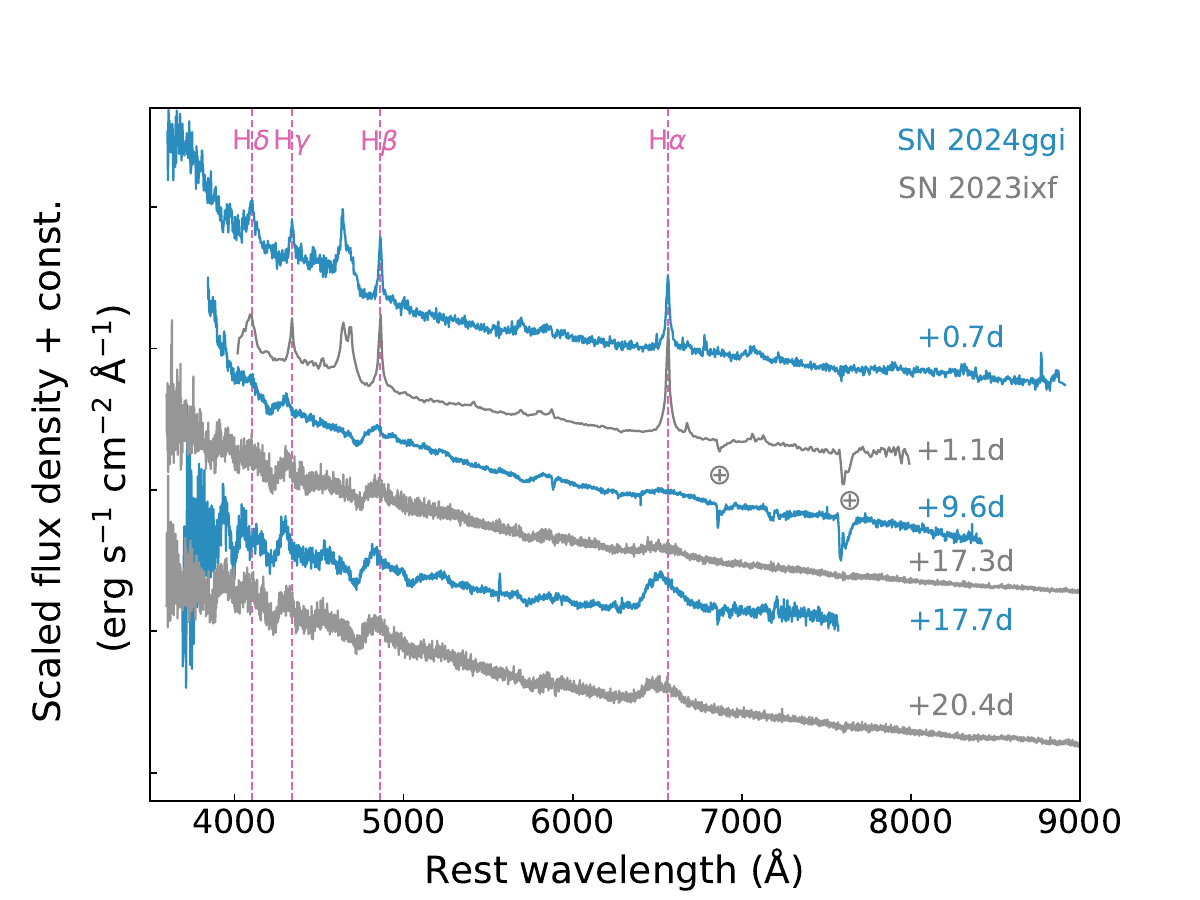}}
    \caption{
        \textit{Upper Left panel:} ATLAS pre-discovery, discovery, and rapid follow-up images of SN~2024ggi are presented, with their corresponding subtracted images (observed MJD shown in red) displayed at the bottom.
        \textit{Bottom Left panel:} A $gri$-colour composite image of SN~2024ggi, its host galaxy, and the environment. This image was created using SLT g-, r-, and i-band (blue, green and red respectively) images taken between 2024-04-11 and 2024-04-20. We homogenized the background flux, combined the images using median stacking, and processed them with \texttt{PixInsight} to enhance colour and contrast. 
        \textit{Upper Right panel:} Multi-band light curves of SN~2024ggi. The citizen science images are provided by Shiau and Man, as well as the Taipei Astronomical Museum and CheCheng Elementary School Observatory. We converted these magnitudes to the r band.
        \textit{Bottom Right panel:} Spectroscopic evolution of SN~2024ggi and comparison with SN~2023ixf. 
    }
    \label{fig:dis_colour}
\end{figure*}

Our ATLAS Transient Server requires three or more spatially coincident detections 
(with significance of at least 5$\sigma$) 
of a source to trigger it as an object for further processing \citep[see][for more details]{Smith2020}. 
The three detections of the source internally labelled as ATLAS24fsk, and automatically 
associated with the nearby galaxy NGC~3621, were made on 11 April 2024, at UT 03:23 (MJD = 60411.141), UT 05:36 (60411.234) and UT 06:01 (60411.251) in 110 sec exposures. The three detections indicated a rapidly brightening transient on the \textit{orange} ($o$) filter, analogous to the Pan-STARRS/SDSS $r + i$ filters. With 
the first detection at $o=18.95\pm0.10$, 
the three detections immediately revealed a rapid intra-night rise of 1.03\,mag over a span of 2.64\,hours. This implies an absolute magnitude of $M_{o}=-10.32$ to $-11.35$ mag, after Milky Way extinction correction 
\citep[$A_{o}=(A_{r}+A_{i})/2=0.16$;][]{Schlafly2011} and adopting a host distance modulus of $\mu=29.11$ at a luminosity distance of 6.64\,Mpc (for further discussion on distance, see Sec.\,\ref{sec:analysis_host}). We immediately posted the discovery on the Transient Name Server\footnote{https://www.wis-tns.org/object/2024ggi}, with the IAU name AT~2024ggi at sky coordinates of RA $= 169.59207$ and Dec $= -32.83759$ ($11^{\rm h} 18^{\rm m} 22.09^{\rm s}$, $-32^\circ 50' 15.3''$) \citep{2024TNSTR1020....1T} on 2024 April 11 09:03 UT. We also posted an AstroNote to draw attention to the discovery \citep{2024TNSAN.100....1S}. We had a gap in observations prior to this mostly due to weather, with the last non-detection six days prior on MJD 60405.063, with an $o$-band limit of 19.8 mag. This is not a constraining limit but the rapid rise within the 2.64\,hrs spanning the exposures indicated this was likely a very young SN. 
Figure\,\ref{fig:dis_colour} shows the discovery image, along with the reference and the subtracted frames. 
We conducted a search for precursors in the history of forced photometry in ATLAS (see Sec.\,\ref{sec:analysis_precursor} for details).

\subsection{Kinder early-phase, high-cadence follow-up}

The Kilonova Finder (Kinder) project is dedicated to rapidly identify fast-evolving transients, especially those displaying blue/red colours and rapid fading characteristics, with the specific aim of detecting kilonovae. 
We use the 40\,cm SLT telescope at Lulin Observatory in Taiwan as the primary instrument for observing newly discovered nearby transients within 100\,Mpc, found by ATLAS, hence exploiting the longitude difference as night hours move west. Equipped with standard SDSS filters ($u'$, $g'$, $r'$, $i'$, and $z'$), the SLT facilitates the efficient selection of objects exhibiting significant colour indices. Moreover, we have developed a dedicated pipeline, known as Kinder-pip \citep{2021A&A...646A..22Y}, to perform image subtractions using archival images sourced from databases such as the SDSS, Pan-STARRS1, and DESI legacy survey. Since its first follow-up campaign in 2021 \citep{2021TNSAN..92....1C}, the Kinder project has investigated over 280 objects, with some being used in detailed single-object studies \citep{2024ApJ...960...29P,2024arXiv240410660G,2024arXiv240513596M}.

During the ATLAS eyeballing process of the initial three detections of SN~2024ggi, we triggered multi-band follow-up imaging observations with SLT in order to confirm this discovery and further constrain the rapidly evolving light curve. Thanks to Lulin Observatory's longitude, we were well placed to rapidly slew to SN~2024ggi following the ATLAS discovery. Observations started at 11:24 UT on 11 of April 2024 (MJD = 60411.476) as soon as the target was visible during evening twilight. We clearly detected SN~2024ggi in the first $u$ and $g$-band raw images and confirmed it as a real source, prompting the ATLAS team to publish the AstroNote \citep{2024TNSAN.100....1S}
describing the discovery. 
Following image reduction and photometric measurements, we identified SN~2024ggi to be a blue and fast-evolving transient: in particular the $r$ band exhibited a sharp rise of 2.56\,mag in 8.26 hours compared to the ATLAS $o$-band discovery. Our results were reported to TNS in \citet{2024TNSAN.102....1C}. Concurrently, the GOTO project also reported a similar magnitude \citep{2024TNSAN.101....1K}. 
We conducted continuous observations of SN~2024ggi using SLT. On the first night the observing conditions were good (with seeing around 1.2 arcseconds) and observations were carried out down to very high airmass (4.42) which allowed 6 hours of continuous coverage of the early rise of the light curve. We obtained a total of 53 frames. The brightness increased by 1.2\,mag in the $u$ band, and by 0.8, 0.7, 0.5, and 0.6 in the $g$, $r$, $i$, and $z$ bands, respectively.

We employed Kinder-pip \citep{2021A&A...646A..22Y} to conduct aperture photometry for SN~2024ggi without template subtraction. Magnitudes were determined by calibrating against SkyMapper field stars. We report the first epoch of each band's magnitude in Table\,\ref{tab:lc_spec_log}, and the complete measurements in a dedicated machine-readable table. 
In addition, we measured the magnitudes using various methods, including aperture and PSF photometry, both with and without template subtraction (against the archival Legacy Survey template images). 
The results are generally consistent, despite a few early points showing a 0.2 magnitude difference. Due to the lack of $u$-band template images, we decided to present and adopt the aperture photometry without template subtraction in this Letter to maintain uniformity. However, for specific cases like searching for shock breakout emissions in the early phases, we used PSF magnitudes after template subtraction instead. Measurements obtained through different methods will be made publicly available in a machine-readable table as well. 
The bottom left panel of Fig.~\ref{fig:dis_colour} displays a colour composite image from SLT images.

\subsection{Citizen science images}
Citizen science images regularly contribute valuable data for discovering and studying SNe, especially during the early phases which are crucial for constraining the explosion time and rise, and investigating SBO phenomena \citep[e.g., SN~2023ixf][]{2023TNSTR1158....1I,2023TNSAN.133....1Y,2024Natur.627..754L}. 
Based on our experience with SN~2023ixf \citep{2023TNSAN.175....1C}, we requested early-phase images of SN~2024ggi via the Facebook Taichung Astronomical Association group. 
Five groups, including citizen astronomers Shiau, Man, and Kuo, as well as CheCheng Elementary School Observatory (CCESO) and Taipei Astronomical Museum (TAM), provided data taken from 2024-04-11.52 to 2024-04-20.58 (Table\,\ref{tab:lc_spec_log}). Unlike M101, NGC~3621 is less popular among citizen scientists and has low-altitude visibility from Taiwan, resulting in no pre-discovery observations.

\subsection{Pan-STARRS1}
In order to supplement our high-cadence SLT photometry, we obtained additional photometric followup with the 1.8m Pan-STARRS1 telescope in Hawaii \citep[PS1;][]{Chambers2016} in the $grizy_{\rm PS}$ filter system \citep{Tonry2012}. PS1 is equipped with a 1.4 Gigapixel camera, with a 0.26 arcsec pixel scale and a 7 square degree field of view. The images were processed using the Image Processing Pipeline \citep[IPP;][]{Magnier2020a}. PS1 $3\pi$ survey data was used as reference for image subtraction and PS1 reference stars in the field were used for zero-point calibration \citep{Magnier2020b}.

\subsection{Neil Gehrels Swift}
SN~2024ggi was also observed with the Ultra-Violet/Optical Telescope \citep[UVOT;][]{Roming2005} on board the Neil Gehrels Swift observatory\footnote{PIs: Sand, Schulze, Hoogendam, Zimmerman, Ravi} \citep{Gehrels2004}. The UVOT photometry was performed using the task \texttt{uvotsource} within HEASoft version 6.25, with a 5 arcsec aperture. We do not perform host subtraction. We note that the SN is saturated in the $u$ band images from MJD 60413 to 60422. 
We do not report the saturated magnitudes or use them for any further analysis in this Letter.

\subsection{Spectroscopic classification and follow-up} 
\label{sec:spec}

\subsubsection{Li-Jiang 2.4 m telescope}
The classification spectrum of SN~2024ggi \citep{2024TNSCR1031....1Z} was obtained at Li-Jiang Observatory of Yunnan Observatories (YNAO) using the Li-Jiang 2.4 m telescope (hereafter LJT; \citealp{2015RAA....15..918F}) equipped with the YFOSC (Yunnan Faint Object Spectrograph and Camera; \citealp{2019RAA....19..149W}) on MJD 60411.608. This spectrum underwent standard reduction procedures in \texttt{IRAF}, encompassing wavelength and flux calibration as well as correction for telluric absorptions. The spectral resolution of this data is estimated to be approximately 460, determined from the Full Width at Half Maximum (FWHM) of skylines.
The spectrum obtained by LJT matches a young SN II with flash features due to SN-CSM interaction (see Fig.\,\ref{fig:dis_colour}).
The redshift estimate from the average of narrow H$\alpha$ and H$\beta$ lines is $z=0.00214$.

\subsubsection{Lulin One-meter Telescope}
Follow-up optical spectra were obtained using the Lulin One-meter Telescope (LOT) with LISA, a commercial spectrograph produced by the Shelyak company, with a resolving power of around 1000 using a 300 line/mm grating. Adopting the primary QSI 660w CCD camera, it provides a wavelength coverage between 3700 and 8436 \AA\ with a pixel resolution of 1.8 \AA. We followed the standard analytic process to subtract the bias and dark and flatten the two-dimensional spectral image with the dome flat illuminated by a tungsten lamp. To perform the wavelength calibration, the in-built ArNe lamps were used. 
We selected spectrophotometric standards observed at a similar elevation to the target to construct the response curve along the wavelength channel, to rebuild the relative intensity of our observation. We used the standard stars HR7596 on 2024-04-20 and HIP~47431 on 2024-04-28, respectively. 
However, we caution that since the observations were of TOO on a moist night, no absolute flux calibration is available.
The spectroscopic observations are detailed in Table\,\ref{tab:lc_spec_log} and the extinction corrected spectra are presented in Fig.\,\ref{fig:dis_colour}. We highlight this is the first coordinated observation for SN follow-up from the Lulin Observatory, combining spectroscopic observations using LOT with simultaneous photometric observations using SLT.

\section{Analysis} \label{sec:analysis}

\subsection{Precursor search for SN~2024ggi}  \label{sec:analysis_precursor}

We employed the ATClean \citep[ATLAS Clean;][]{2024arXiv240503747R} package to search for possible precursors in difference images from the ATLAS survey. ATClean forces photometry at the position of transient in historical ATLAS data (in the difference images) and analyses the results by cleaning the individual measurements using statistical methods and the flux, uncertainty and point-spread-function measurements on the detector. Systematic residuals in the images that arise from detector artifacts, nearby sources or imperfect image subtraction can all mimic an astrophysical flux excess. ATClean uses control light curves around a source, which involves forcing multiple photometric measurements close to the source, and comparing the 
statistical significance of the photometry forced at the source with that
in the control fields. These control light curves are utilized to clean the photometry and calculate more robust detection limits for potential pre-SN eruptions than just forcing at the position of SN~2024ggi alone. With this method, we do not find any evidence for 
any significant real, astrophysical, excess flux in the historical ATLAS 6.5 year data between MJD = 58065 (2017-11-08) and 60405 (2024-04-05).
\cite{2024arXiv240503747R} describe how we can estimate efficiency of recovery of simulated precursor with certain flux and duration, using a Gaussian profile for the flux of a precursor event. The longer the duration of the precursor, then the larger the time window we can bin over, allowing us to probe fainter fluxes for longer duration simulated
precursors. We have 90\% detection efficiency for simulated Gaussian eruptions with $\sigma_{sim}\ge 80$ days with peaks $o=20.5$ mag.
For shorter duration events, we have shallower detection limits of 90\% efficiency at  17.5, 19.5, 20 mag for $\sigma_{sim}$=5, 20, and 40 days, respectively.
Our result is consistent with \citet{2024arXiv240518490S}, they do not detect any precursor emission for SN~2024ggi down to $-9$\,mag.

\subsection{Shock breakout signal search}
\label{sec:analysis_explosion_sbo}

Constraining the explosion epoch is crucial to calculating accurate rise times and other light-curve fitting endeavours. 
As discussed in Appendix \ref{sec:analysis_explosion}, we modeled the early rise of light curves for SN~2024ggi using power law fits similar to those in \cite{2020ApJ...902...47M}, and determined the explosion epoch to be MJD $60410.90^{+0.08}_{-0.12}$, which will be used throughout the Letter.

The power law fits were also used to search for potential SBO emissions in the early light curves of SN 2024ggi, similar to what was found in the early data of SN 2023ixf.
As shown in the lower panel of Fig. \ref{fig:powerlaw}, we present our multiband light curves collected within four days after the explosion. A power law fit was applied to the early phases (i.e., the first $\sim1$ day), and no clear variances between the data and fits were observed, indicating that no apparent SBO features appeared.
One might argue that the absence of SBO is due to observational limits; however, it could also indicate the presence of a dense, extended circumstellar material (CSM) around the red supergiant (RSG). This scenario is further supported by other analyses, such as early colours, as discussed throughout the letter.

\begin{figure*}
    \centering
    \subfigure{\includegraphics[width=0.8\textwidth]{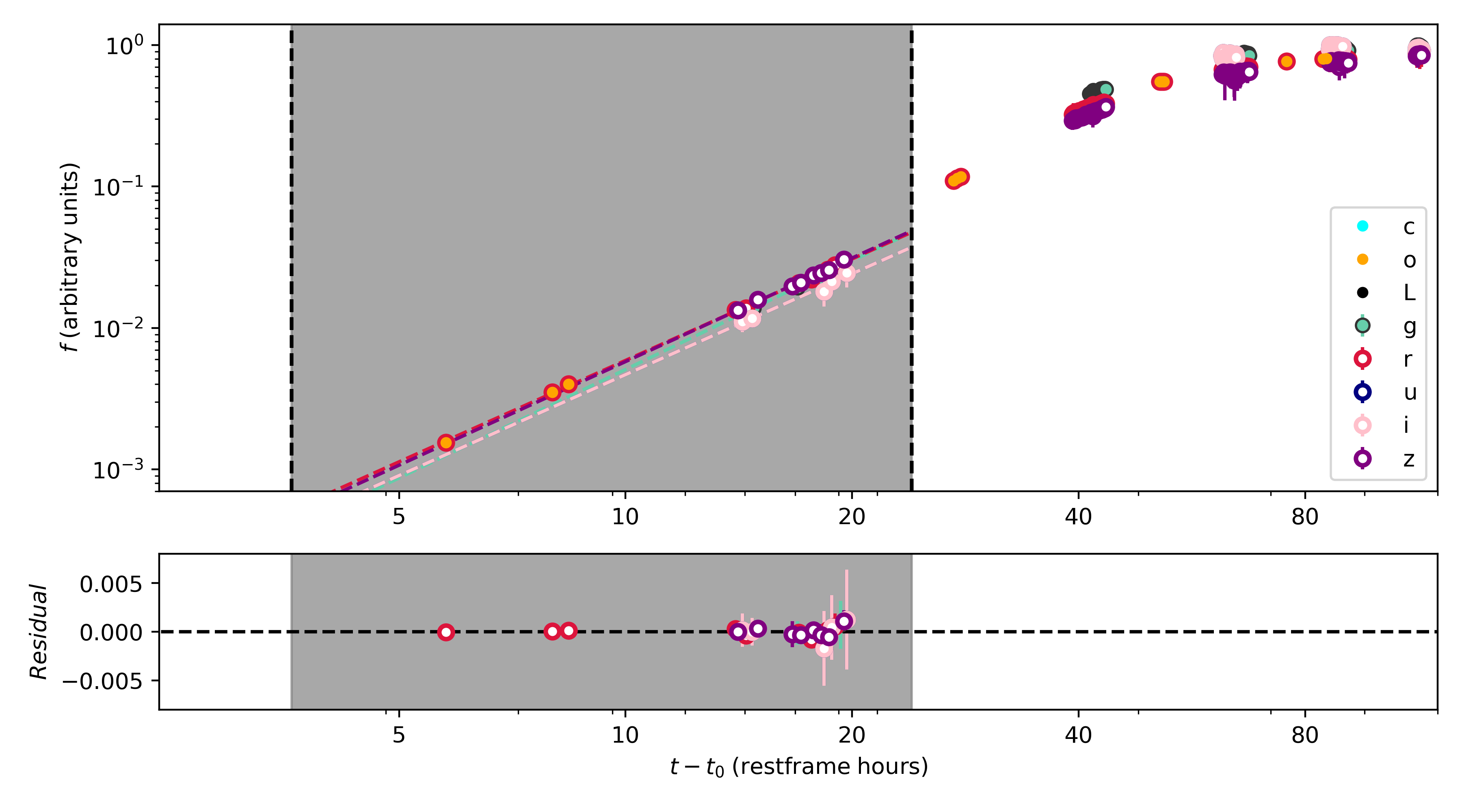}}
    \caption{
        This figure presents the bestfit broken power-law fit to the early optical light curves in order to check the existence of shock cooling breakouts. The vertical gray-shaded area indicates the time interval adopted for the light-curve fitting. The residuals are shown in the lower panel. The error bars shown represent 1 $\sigma$ uncertainties of magnitudes.
    }
    \label{fig:powerlaw}
\end{figure*}

\subsection{Photometric behaviours and colour evolution}
\label{sec:analysis_phot_colour}

To estimate the photometric behaviors and align observations from different bands for spectral energy distribution (SED) construction, such as calculating colours, performing blackbody fits, etc., we interpolate our multi-band light curves using HAFFET \citep{2023ApJS..269...40Y}. 
Given that 2024ggi clearly exhibits characteristics of a Type IIP SN, we opt to utilize the analytical function outlined in \cite{Villar2019}. 
With the interpolated light curves, we found that the plateau occurred at $r=12.02\pm0.09$\,mag (after extinction correction) after 3.6 rest-frame days post-explosion. In other bands, the photometric behavior followed a similar trend until $\sim18$ days post-explosion, except for $UVM2$ and $UVW2$, which exhibited slight declines instead of plateauing. It is evident that SN~2024ggi experienced a rapid rise in the first few days, with a rise of 6.78\,mag ($r$ band) in 3.6 days or less, indicating a slope greater than 1.88\,mag per day. Following this initial rise, a plateau phase of $18+$ days (up to the current time) ensued, firmly establishing SN~2024ggi as a Type IIP SN. However, the plateau is not entirely flat. According to our analytic model fitting, there is an obvious decline that levels off in the UV bands i.e., Swift $u$, $UVM2$, and $UVW2$ bands, while in the red filters, i.e., $r$ and $i$ bands, the light curve plateau is relatively flat. We are only able to see these subtleties thanks to our well-sampled light curve.

\begin{figure*}
    \centering
    \subfigure{\includegraphics[width=0.5\linewidth]{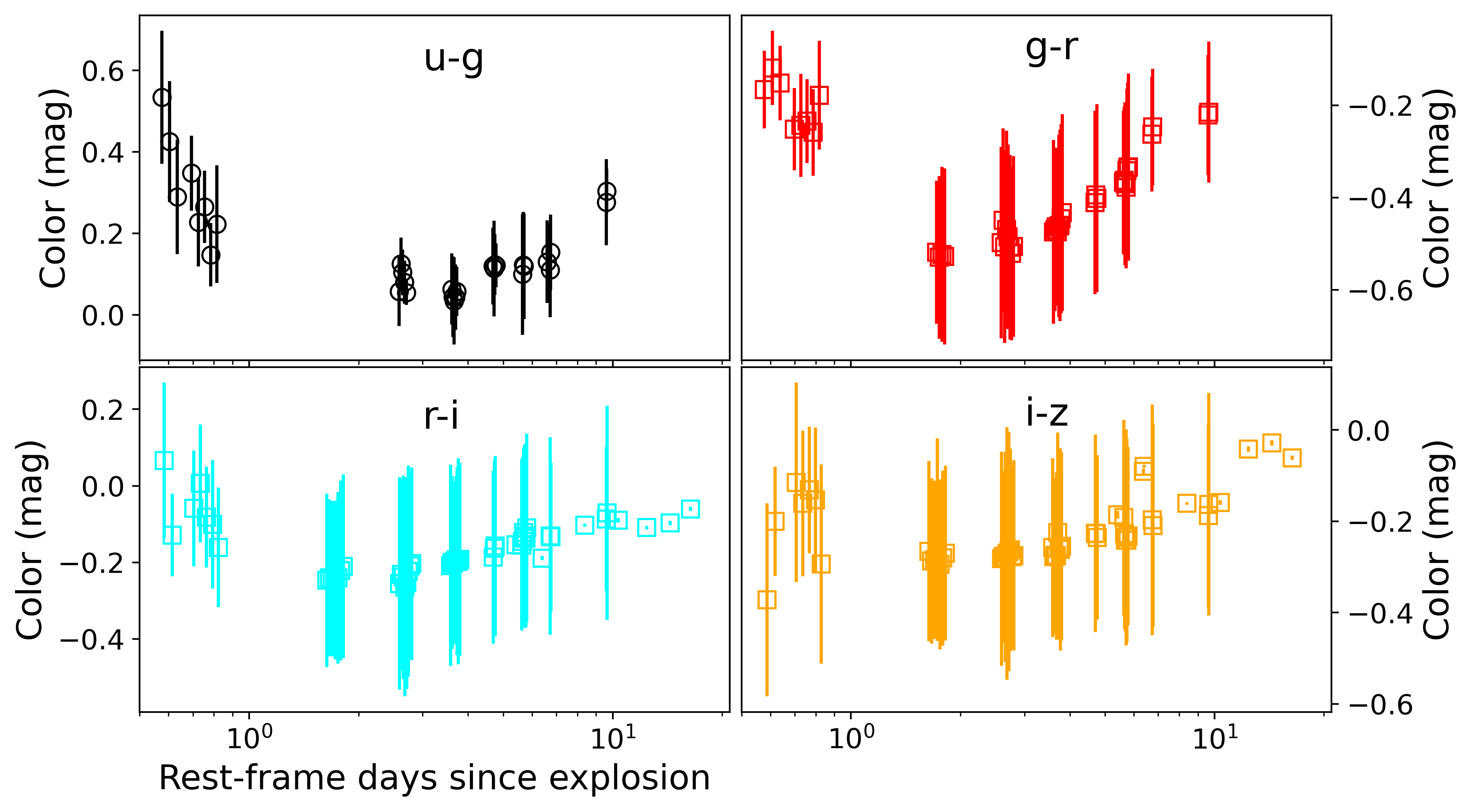}}
    \subfigure{\includegraphics[width=0.45\textwidth]{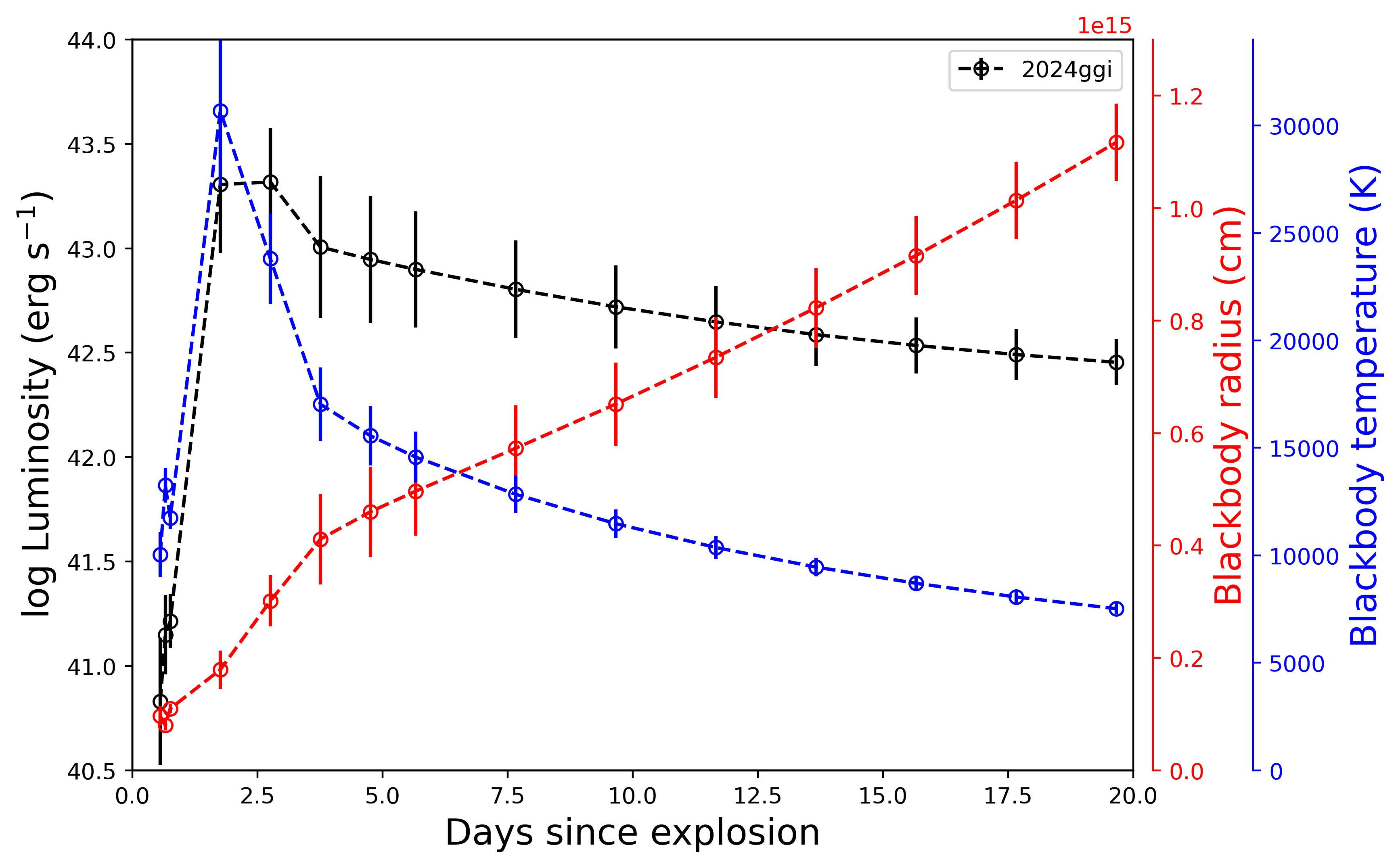}}
    \caption{
    \textit{Left panel:} The early-phase optical colour evolution of SN~2024ggi, including $u-g$, $g-r$, $r-i$, and $i-z$. 
    \textit{Right panel:} Blackbody inferred bolometric light curve (black), temperature (blue) and radius (red) evolution of SN~2024ggi.
    }
    \label{fig:colour_bb}
\end{figure*}

In the left panel of Fig. \ref{fig:colour_bb}, we investigate the optical colour evolution of SN~2024ggi. All the colours were obtained by aligning the photometry in different bands using a 1-day bin. As shown, while other bands generally remain constant, the $u-g$ colour decreases rapidly from 0.5 to 0\,mag within just 6\,hours, observed $\sim0.57$\,day after the explosion.
All of these findings support the scenario of a dense, extended CSM surrounding the RSG progenitor.

\subsection{Bolometric light curve, temperature and radius evolution}
\label{sec:analysis_bol_temp_ra}

We construct a bolometric light curve using the multiband photometry obtained for SN 2024ggi. 
To ensure that the fluxes are based more on observational data rather than solely on predictions, we matched photometric epochs using a 1-day bin and selected those epochs with observations in more than three bands. The remaining wavelength fluxes were estimated using analytic models outlined in \cite{Villar2019}, as described in Sec. \ref{sec:analysis_phot_colour}.
Subsequently, we fit a blackbody spectrum to each of these single-epoch SEDs using methods similar to Superbol \citep{superbol} to derive the bolometric luminosity, temperature, and photospheric radius, shown in the right panel of Fig. \ref{fig:colour_bb}. As illustrated, we find the temperature at 0.6 days to be 10,000\,K, peaking at 31,000\,K at $+1.7$ days post explosion epoch. These temperatures are consistent with those found by \cite{2024arXiv240419006J} and \cite{2024arXiv240518490S}, and slightly higher at the peak than those reported by \cite{2024arXiv240507964C}, who did not include UV data for temperature calculations. The peak temperature for SN~2024ggi is close to that observed for SN~2023ixf by \cite{Zimmerman2024}. The implied radius at our first epoch, at 0.6 days post-explosion, is $\sim0.8\times10^{14}$\,cm ($\sim$1149 $R_\odot$), which is smaller than the radius calculated for SN~2023ixf of $\sim1.9\times10^{14}$\,cm ($\sim$2731 $R_\odot$) by \cite{Zimmerman2024}.
Assuming an initial progenitor radius of 600\,$R_\odot$, this suggests that the photosphere has expanded outward at a velocity of only 7500\,km\,s$^{-1}$, which is relatively slow for less than a day after explosion. This may indicate a significant amount of mass overlaying the progenitor, or that the photosphere is also receding inwards in Lagrangian coordinates as the photospheric temperature rapidly cools.

\subsection{Spectroscopic properties}
\label{sec:analysis_spec}

Our $+0.7$\,day classification spectrum of SN~2024ggi shows narrow emissions of high-ionization species, known as flash features, as in Fig.\,\ref{fig:dis_colour}. Comparing with the $+1.36$\,day spectrum of SN~2023ixf and lines identified from \citet{2023ApJ...956L...5B,2023SciBu..68.2548Z}, we detected the Balmer series (H$\alpha$, H$\beta$, H$\gamma$, H$\delta$), 
He~\I\ $\lambda 6678.15$, $\lambda 7065.19$, 
C~\III\ (possible $\lambda 4056.0$), $\lambda\lambda 4647.5, 4650.0$, $\lambda 5695.9$, 
N~\III\ $\lambda 4097.33$, $\lambda\lambda 4634.0, 4640.64$, $\lambda 4858.82$ (blended with H$\beta$), 
and possible N~\IV\ lines blended with the He~\I\ line.
We did not detect the O~\III\ lines seen in SN~2023ixf.

We compare the spectroscopic evolution of SN~2024ggi with the nearby SNe~IIP 2023ixf that also exhibits flash features\footnote{The comparison spectra were downloaded from WiseREP \citep{2012PASP..124..668Y} for SN~2023ixf \citep[][and DESI]{2023TNSCR1164....1P}. All spectra were calibrated for Milky Way and host extinction, with phases relative to the explosion time. We used $E(B-V){\rm total} = 0.04$ mag and MJD 60082.788 for SN~2023ixf \citep{2024Natur.627..754L}}.
SN~2024ggi shows significant similarity to SN~2023ixf. Flash features are present in the first-day spectrum but disappear by $+9.6$\,days, leaving a blue, featureless spectrum with emerging H$\beta$. By $+17.7$\,days, the P-Cygni profiles of H$\beta$, H$\gamma$, and H$\delta$ are visible, though H$\alpha$ absorption component is not yet present. SN~2023ixf also lacks this absorption at a similar epoch. 
In the $+17.7$\,day spectrum of SN~2024ggi, all Balmer P-Cygni profiles have blueshifted emission peaks. We measured a velocity of $-1840$\,km\,s$^{-1}$ from the H$\alpha$ emission. \citet{2014MNRAS.441..671A} systematically studied this phenomenon and concluded that blueshifted emission-peak offsets are a generic property of photospheric-phase Type II SNe.

\section{Discussions} 
\label{sec:discussions}

\begin{figure*}[ht!]
    \centering
    \subfigure{\includegraphics[width=0.48\textwidth]{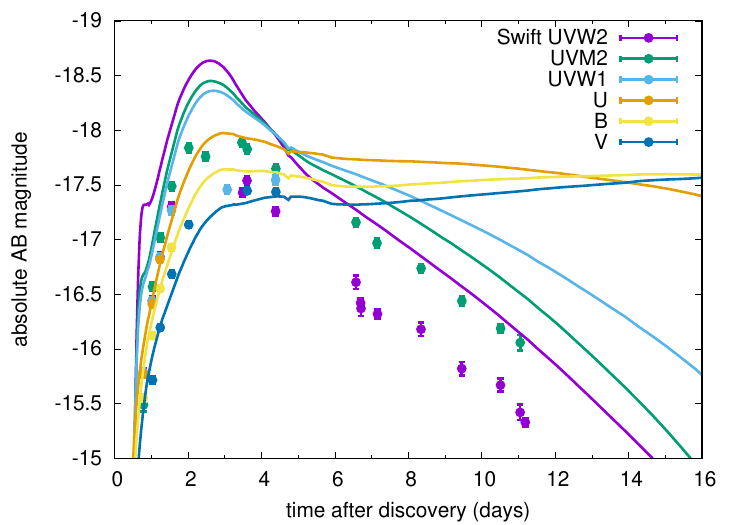}}
    \subfigure{\includegraphics[width=0.48\textwidth]{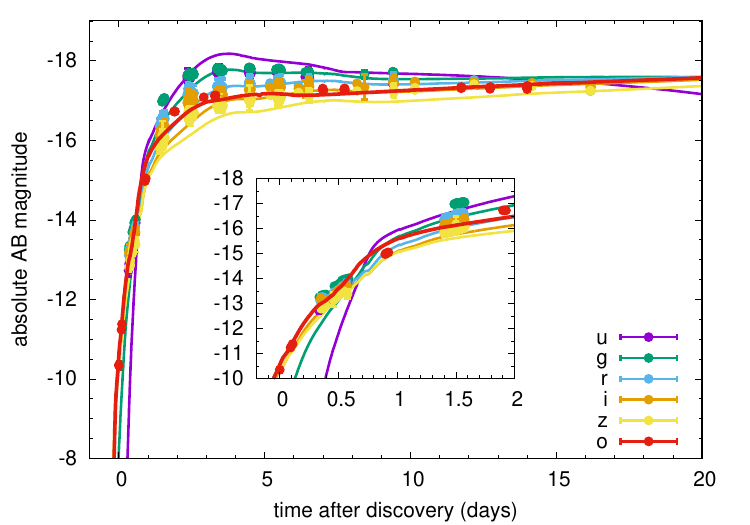}}
    \caption{
        \textit{Left panel:} Light curve modelling for SN~2024ggi in the UV wavelength. 
        \textit{Right panel:} Light curve modelling for SN~2024ggi in the optical bands. 
    }
    \label{fig:LC_modelling}
\end{figure*}

\subsection{Modelling results}
To constrain the confined CSM properties based on the early phase light curves, we searched for the best-matching models from the pre-computed light-curve library presented in \citet{2023PASJ...75..634M}. The light-curve library contains multi-frequency light curves from the explosions of the solar-metallicity RSG SN progenitors with the ZAMS mass of $10-18~\mathrm{M_\odot}$ computed by \citet{2016ApJ...821...38S}. The light curves are computed by the one-dimensional radiation hydrodynamics code \texttt{STELLA} \citep{1998ApJ...496..454B,2000ApJ...532.1132B,2006A&A...453..229B}. The model library covers explosion energies of $0.5-5\times 10^{51}~\mathrm{erg}$ and $^{56}\mathrm{Ni}$ masses of $0.001-0.3~\mathrm{M_\odot}$. All the models are assumed to have a confined CSM. The mass-loss rates and radii of the confined CSM are assumed to be in the ranges of $10^{-5}-10^{-1}~\mathrm{M_\odot~yr^{-1}}$ and $10^{14}-10^{15}~\mathrm{cm}$, respectively. The terminal wind velocity is assumed to be $10~\mathrm{km~s^{-1}}$ with the wind acceleration parameters $\beta$ in the range of $0.5-5$. We refer to \citet{2023PASJ...75..634M} for further details of the model library.

Because we focus on the early phase in the light curves that are mainly affected by the dense confined CSM, we do not constrain the $^{56}\mathrm{Ni}$ mass in this Letter. Similarly, because the early interaction signature is dominated by the CSM interaction, the progenitor mass is not well constrained, although we did not fix the progenitor mass in searching for matching light-curve models. We searched for a model that matches the optical luminosity evolution well. The best matching model we found is presented in Fig.\,\ref{fig:LC_modelling}. The model has a explosion energy of $2\times 10^{51}~\mathrm{erg}$, a mass-loss rate of $10^{-3}~\mathrm{M_\odot~yr^{-1}}$ (with the assumed $10~\mathrm{km~s^{-1}}$ wind), a confined CSM radius of $6\times 10^{14}~\mathrm{cm}$, and $\beta =4.0$. The corresponding CSM mass is $0.4~\mathrm{M_\odot}$.  
The previous mass-loss rate estimate by \citet{2024arXiv240419006J} is slightly higher ($10^{-2}~\mathrm{M_\odot~yr^{-1}}$ with $50~\mathrm{km~^{-1}}$), but their estimated explosion energy ($1.2\times 10^{51}~\mathrm{erg}$) and confined CSM radius ($4\times 10^{14}~\mathrm{cm}$) are slightly lower.
This model has a progenitor ZAMS mass of $12~\mathrm{M_\odot}$, but the ZAMS mass is difficult to constrain in this phase.

Our model has the higher UV luminosity than observed. If we try to match the UV luminosity, the optical luminosity becomes lower than observed and we did not find a good model matching both optical and UV. Similarly, the light-curve model presented in \citet{2024arXiv240419006J} matches well in UV, but their optical luminosity is fainter than observed. These discrepancies may originate from uncertainties in extinction estimates as well as the assumption of spherical symmetry in both models.

Using the same model grid, \citet{2024arXiv240520989S,2024arXiv240600928M} estimated the properties of SN~2023ixf. The mass-loss rate is estimated to be around $10^{-2}~\mathrm{M_\odot~yr^{-1}}$ with $10~\mathrm{km~^{-1}}$ with the confined CSM radius of around $6\times 10^{14}~\mathrm{cm}$ with the explosion energy of $2\times 10^{51}~\mathrm{erg}$. The CSM mass is $0.85~\mathrm{M_\odot}$. The estimated mass-loss rate and thus CSM mass for SN~2024ggi are slightly lower than those of SN~2023ixf.

We consider solar metallicity for SN~2024ggi in our model assumption based on host galaxy studies. 
Investigations by \citet[][]{2018MNRAS.480.1973K} indicate NGC\,3621 to be already very metal rich in its early time. While studying the circumstellar environment around the progenitor, \citet[][]{2024arXiv240419006J} have indicated the presence of solar metallicity CSM. The analysis by \citet[][]{xiang2024red} indicate the progenitor star to be the reddest and distinct among the red stars in the nearby vicinity, which too leads them to assuming solar metallicity. Further, \citet[][]{chen2024early} also fit the solar metallicity isochrones to estimate the constraints over the range of progenitor mass. 


\subsection{Comparison with other SNe~II}

\begin{figure*}[ht!]
    \centering
    \subfigure{\includegraphics[width=.9\textwidth]{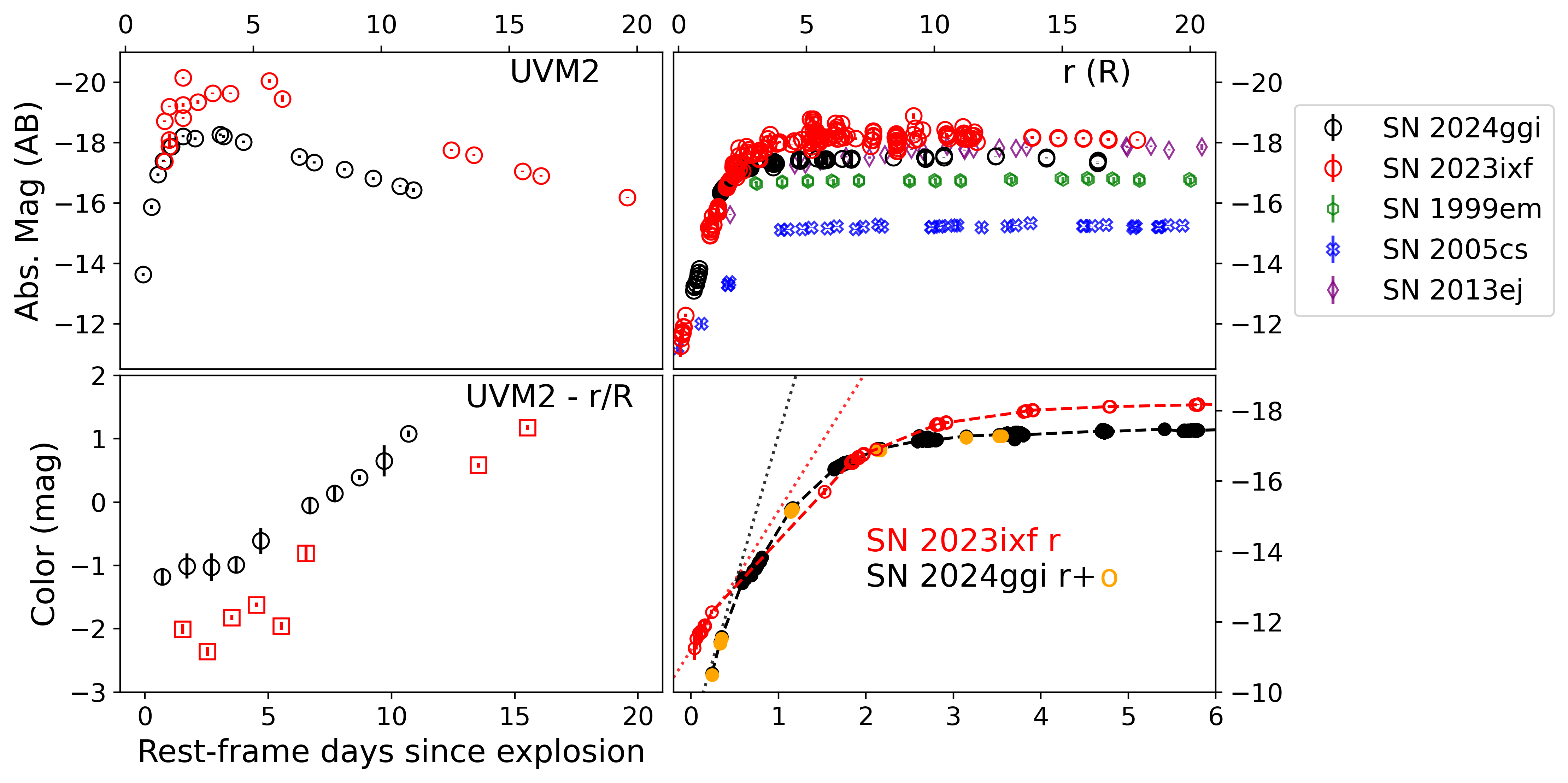}}
    \caption{
        UVM2 and r/R band Light curve, as well as their colours, comparison of SN~2024ggi, SN~2023ixf and other Type IIP SNe. \textit{Upper left panel:}  
        \textit{Upper Left panel:} $UVM2$ band light curves.  
        \textit{Upper right panel:} $r/R$ band light curves.
        \textit{Bottom left panel:} $UVM2-r/R$ colours. 
        \textit{Bottom right panel:} A zoom-in view of the $r$-band light curves, with the adopted Villar analytic model depicted by dashed lines and linear fits of the early rise indicated by dotted lines.
    }
    \label{fig:LC_comp}
\end{figure*}

We selected well-observed and well-understood typical Type II SNe as a comparison sample, covering a wide range of luminosities (SNe~2013ej, 1999em, 2005cs, in order from bright to faint). This sample also includes SNe with spectra that exhibit flash features (SN~2023ixf).
The upper panels in Fig.\,\ref{fig:LC_comp} present the $UVM2$ and $r/R$-band light curves, along with their colours in the bottom left panel. 
SN~2024ggi is slightly fainter in the $r$ band compared to SN~2023ixf, with $-17.8$\,mag at the plateau. SN~2024ggi is significantly dimmer in the UV wavelength comparing with other flashers, as reflected in the $UVM2-r$ colour. Relative to standard SNe~II, early-time flashers are generally very bright in the UV \citep{2023arXiv231016885I,2024arXiv240302382J}.
Another noteworthy phenomenon is that SN~2024ggi exhibited a much faster rise (black dotted line) compared to SN~2023ixf (red dotted line) in the bottom right panel.

\section{Conclusion and future prospect}
\label{sec:conclusion}

The ATLAS early (six hours after the SN explosion) detection and subsequent high-cadence monitoring with Kinder project of SN~2024ggi (in between 14 and 20 hours after the explosion) have provided a wealth of data that is invaluable for understanding the early stages of SN evolution. All these photometric (e.g. getting bluer rapidly) and spectroscopic behaviours indicate a wind SBO scenario. 

The blackbody fitting indicates a radius of $\sim0.8\times10^{14}$\,cm at 0.6 days post-explosion, suggesting a photospheric expansion velocity of 7500\,km\,s$^{-1}$. This slow expansion may imply a significant amount of overlying mass on the progenitor or an inward recession of the photosphere in Lagrangian coordinates as the temperature rapidly cools.

The radiation time of the SBO is related to the progenitor's size, and a dense CSM can make the SBO brighter and longer-lasting. Therefore, comprehensive coverage is crucial for capturing the SBO and studying a diversity of the SN progenitors. Statistics from the ZTF bright transient survey \citep{2020ApJ...895...32F,2020ApJ...904...35P} indicate that, on average, one SN peaks brighter than 15 magnitudes every two months. These bright SNe are ideal targets for citizen astronomers. High cadence, continuous follow-up is essential to capture the SBO signal. Achieving this is impossible at a single site, but it is an ideal project for a global observational network coordinating with citizen scientists. Such a strategy was applied in the case of SN~2023ixf \citep{2024Natur.627..754L}. Now, with SN~2024ggi, we provide another example that demonstrates citizen scientists responded rapidly to followup and the benefits of collaborative high-cadence observations.

\section{Acknowledgments}

We thank Mr. Chi-Sheng Lin and Mr. Jhen-Kuei Guo for Kinder observations and IT support, and thank to citizen astronomer Mr. Wen-Li Kuo for providing the SN JPG images.
T.-W.C. thanks Mrs. Yi-Zhen Lin for her administrative assistance to the GREAT Lab.
K.K.M. thanks Mr. David Cheng, one of his partners of Gemini Remote Observatory, for his feverish effort documenting the supernova despite of far from ideal weather condition. 
T.-W.C. \& A.A. acknowledge the Yushan Young Fellow Program by the Ministry of Education, Taiwan for the financial support. 
S.Y. \& Z.-N.W. are supported by the National Natural Science Foundation of China under Grant No. 12303046 and the Henan Province High-Level Talent International Training Program.
Numerical computations were in part carried out on PC cluster at the Center for Computational Astrophysics, National Astronomical Observatory of Japan.
This work was supported by JSPS Core-to-Core Program (grant number: JPJSCCA20210003).
This work was funded by ANID, Millennium Science Initiative, ICN12\_009.
M.F. is supported by a Royal Society - Science Foundation Ireland University Research Fellowship.
M.N. is supported by the European Research Council (ERC) under the European Union's Horizon 2020 research and innovation programme (grant agreement No.~948381) and by UK Space Agency Grant No.~ST/Y000692/1.
G.P. acknowledges support from ANID through Millennium Science Initiative Programs ICN12 009.
HFS is supported by the Eric and Wendy Schmidt AI in Science Fellowship.
X.Wang is supported by the National Natural Science Foundation of China (NSFC grants 12288102 and 1203300), and the Tencent Xplorer Prize.
J.Z. is supported by the National Key R\&D Program of China with No. 2021YFA1600404, the National Natural Science Foundation of China (12173082), the Yunnan Province Foundation (202201AT070069), the Top-notch Young Talents Program of Yunnan Province, the Light of West China Program provided by the Chinese Academy of Sciences, the International Centre of Supernovae, Yunnan Key Laboratory (No. 202302AN360001).
%
%
We acknowledge WISeREP - https://www.wiserep.org. 
%
This work has made use of data from the Asteroid Terrestrial-impact Last Alert System (ATLAS) project. The Asteroid Terrestrial-impact Last Alert System (ATLAS) project is primarily funded to search for near earth asteroids through NASA grants NN12AR55G, 80NSSC18K0284, and 80NSSC18K1575; byproducts of the NEO search include images and catalogs from the survey area. This work was partially funded by Kepler/K2 grant J1944/80NSSC19K0112 and HST GO-15889, and STFC grants ST/T000198/1 and ST/S006109/1. The ATLAS science products have been made possible through the contributions of the University of Hawaii Institute for Astronomy, the Queen's University Belfast, the Space Telescope Science Institute, the South African Astronomical Observatory, and The Millennium Institute of Astrophysics (MAS), Chile.
The Pan-STARRS1 Surveys (PS1) and the PS1 public science archive have been made possible through contributions by the Institute for Astronomy, the University of Hawaii, the Pan-STARRS Project Office, the Max-Planck Society and its participating institutes, the Max Planck Institute for Astronomy, Heidelberg and the Max Planck Institute for Extraterrestrial Physics, Garching, The Johns Hopkins University, Durham University, the University of Edinburgh, the Queen's University Belfast, the Harvard-Smithsonian Center for Astrophysics, the Las Cumbres Observatory Global Telescope Network Incorporated, the National Central University of Taiwan, the Space Telescope Science Institute, the National Aeronautics and Space Administration under Grant No. NNX08AR22G issued through the Planetary Science Division of the NASA Science Mission Directorate, the National Science Foundation Grant No. AST-1238877, the University of Maryland, Eotvos Lorand University (ELTE), the Los Alamos National Laboratory, and the Gordon and Betty Moore Foundation.
This research is based on observations made with the mission, obtained from the MAST data archive at the Space Telescope Science Institute, which is operated by the Association of Universities for Research in Astronomy, Inc., under NASA contract NAS 5-26555.

\vspace{5mm}
\facilities{ATLAS, Swift, PS1, LO:1m}

\software{ChatGPT\footnote{ChatGPT serves as grammar checker and paraphrasing tool}, IRAF\citep{iraf1,iraf2}, astropy\citep{2013A&A...558A..33A,2018AJ....156..123A}, HAFFET\citep{2023ApJS..269...40Y}, scipy\citep{2020SciPy-NMeth}, numpy\citep{harris2020array}, Source Extractor \citep{1996A&AS..117..393B}, hotpants\citep{hotpants}, matplotlib\citep{Hunter:2007}, emcee\citep{2013PASP..125..306F}}

\appendix

\section{Observational log}
We list our observational log for photometric and spectroscopic follow up in Table\,\ref{tab:lc_spec_log}. We only list the first epoch of each band, the full table with machine readable format is available online.

\input{Tables/SLT_phot}

\section{Power law fits for the Explosion epoch}
\label{sec:analysis_explosion}

\cite{2024arXiv240419006J} fit the bolometric light curve of SN~2024ggi to a suite of hydrodynamical models, constraining the time of first light to MJD = $60410.56^{+0.07}_{-0.12}$. In this letter, we use the Bayesian framework developed by \cite{2020ApJ...902...47M} to model the early rise of light curves simultaneously in multiple bands as a power law. This approach assumes that the epoch of first light is the same across different bands, which is reasonable given the observation cadence and the similarity of SN ejecta opacity at these wavelengths and we apply this framework to the g and r bands simultaneously.
To establish the flux light curve baselines and increase observation cadence, we included ATLAS pre and post discovery observations, assuming similar filter transmission between the SDSS $r$ ($g$) and ATLAS $o$ ($c$) bands. We note that a non-detection was reported by the GOTO team in their $L$ band, which can effectively constrain the flux baselines as well. Given that the $L$ band has a broad transmission covering both $g$ and $r$, we could include it as a baseline point for both bands. 
We employed a Markov Chain Monte Carlo approach via emcee \footnote{emcee.readthedocs.io} to derive the fitted power law models, testing various methods (fitting from the flux baseline up to 40\% or 60\% of the maximum, both with and without the GOTO $L$ band limits).
Using all the aforementioned assumptions and the HAFFET tool \citep{2023ApJS..269...40Y}, we characterize the early emission of SN~2024ggi as shown in Fig. \ref{fig:powerlaw}. Its left panel displays the early light curves of SN~2024ggi along with the best-fit power law models, while the right panel shows the converged Monte Carlo samples as contours. For this analysis, we used the explosion epoch outlined in \cite{2024arXiv240419006J} as our initial reference. Our subsequent power law fits confirmed this explosion epoch, revealing only a 0.34-day offset. As a result, we constrained the explosion epoch to MJD $60410.90^{+0.08}_{-0.12}$, which will be used throughout the letter. Meanwhile, the power law fits yielded indices of $2.68^{+0.83}_{-0.65}$ and $2.38^{+0.58}_{-0.39}$ for the $g$ and $r$ bands, respectively, indicating a faster rise for SN~2024ggi compared to typical SNe~IIP.

\begin{figure*}
    \centering
    \subfigure{\includegraphics[width=0.5\textwidth]{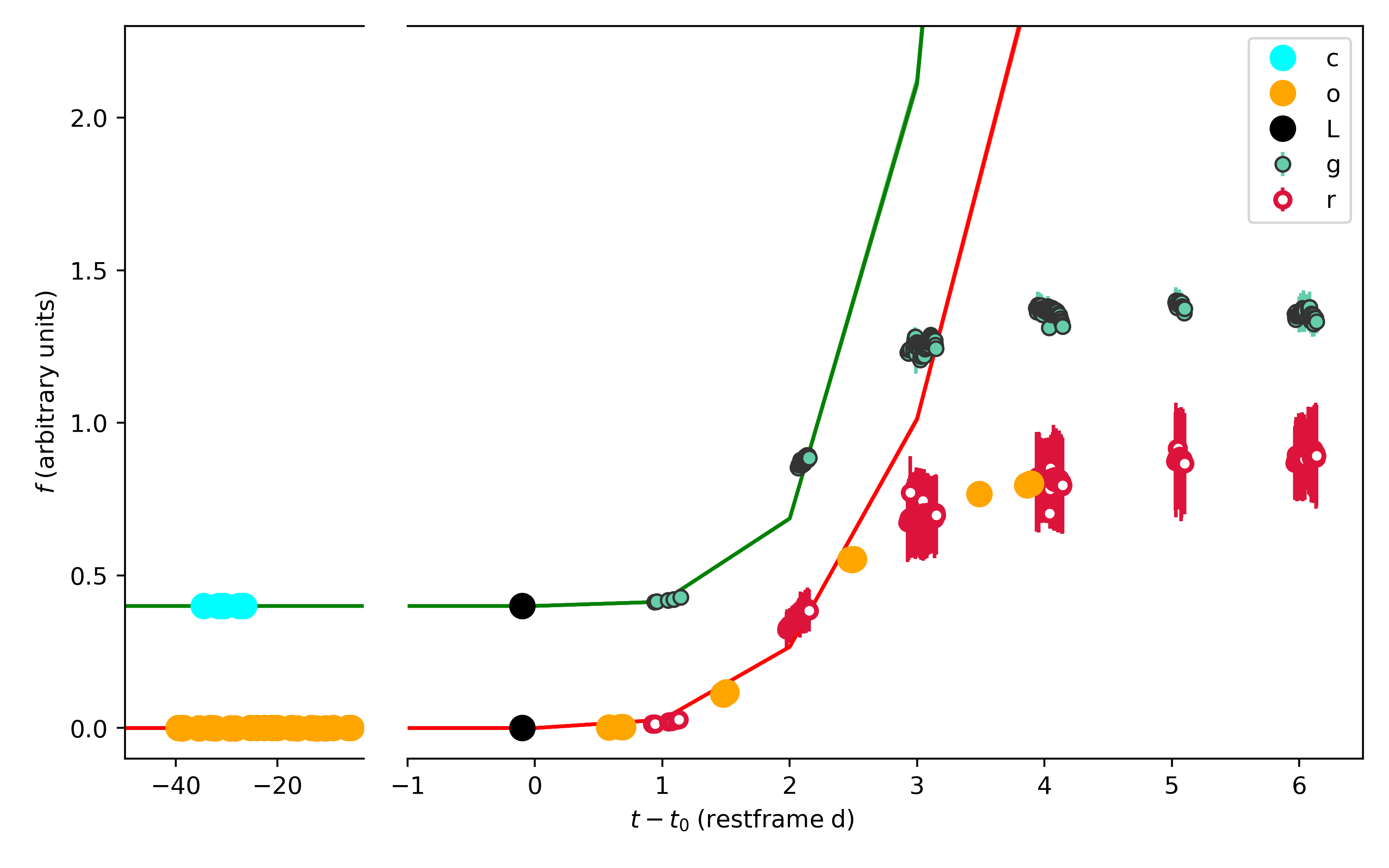}}
    \subfigure{\includegraphics[width=0.45\textwidth]{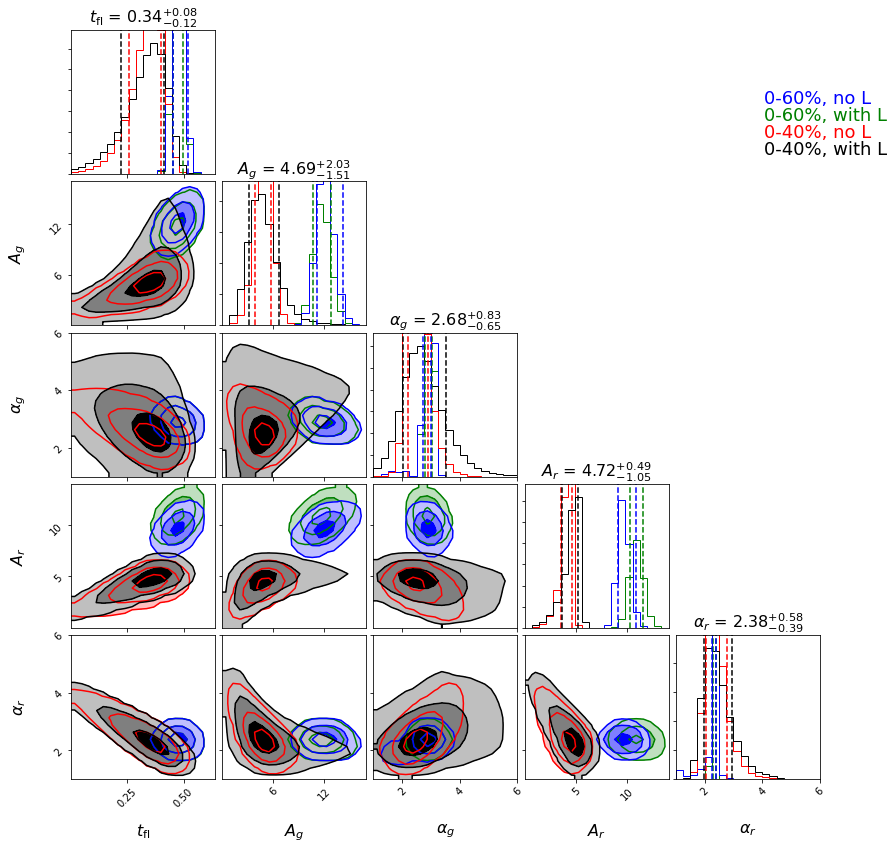}}
    \caption{
        \textit{Left panel:} Power law fits to the early $g$ ($c$) and $r$ ($o$) band light curves, in order to estimate the explosion epoch as well as rising profiles, e.g. power-law indices.
        \textit{Right panel:} Contour plots for the power law fitting parameters are presented, with different colours representing various fitting methods. These methods include fitting from the flux baseline up to 40\% or 60\% of the maximum, both with and without the GOTO $L$ band limits. For each method, the converged parameters are shown within three dashed regions corresponding to 1, 2, and 3 sigma levels, respectively.
    }
    \label{fig:powerlaw}
\end{figure*}

\section{citizen science images calibration}
\label{app:citizen_image_calibration}

The citizen science images on this work were taken by four different sets of instruments; 1. {\bf TAM}: a 30cm reflector with CCD and $L$, $G$, $B$, and Bessel-$I$ filters at Lulin observatory operated by Taipei Astronomical Museum (TAM). 2. {\bf EQMOD}: a 13cm refactor with CCD and $L$ filter own and operated by Lawrence, Dickson, Joe, Paul \& David. The images are provided by Ka Kit Man.
3. {\bf iTelescope 33}: a 32cm f/9 reflector with CCD and $G$, $R$, and $B$ filters at Siding Spring Observatory, Australia. The images are provided by Yu-Chien Shiau.
4. {\bf CCESO}: a 43cm reflector with CCD and $G$, $R$, and $B$ filters at CheCheng Elementary School Observatory (CCESO), Taiwan. 
While the LRGB fitler sets of the four telescopes are not exactly the same, they are mostly consist with: $L$ $\sim$ SDSS $g'$ + $r'$, which has transparency from 400nm to 700nm, and the $RGB$ filters roughly equally divide the wavelength coverage of SDSS $g'$ + $r'$.  
All of the citizen science images were reduced under the standard CCD reduction procedure, which include the bias and dark current subtraction and flat calibration. After removing the instrumental trends, all of the images were calibrated and measured the photometry zeropoints against SkyMapper photometry system \citep{skymapper} by cross matching the on-field sources with Gaia star catalogs and SkyMapper source catalogs. We found the colour conversion between our citizen science instruments to SkyMapper photometry system by solving one of the following equations:
\begin{equation}
m_{citi} - m_{sm} = C_0 + C_1~(g-r)_{sm}, 
\end{equation}
or, 
\begin{equation}
m_{citi} - m_{sm} = C_0 + C_1~(g-r)_{sm} + C_2~(g-r)^2_{sm}
\end{equation}
Here $m_{citi}$ and $m_{sm}$ are the citizen science images and SkyMapper magnitude, respectively, $C_0$ is a constant $C_1$ is the coefficient of linear colour-term, and $C_2$ is coefficient of the quadratic term. 
For the TAM system, we found:
\begin{align*} 
\begin{split}
L - r_{sm} &= -0.20 + 0.63~(g-r)_{sm} - 0.19~(g-r)^2_{sm},~rms=0.05 \\
G - r_{sm} &= -0.27 + 0.72~(g-r)_{sm},~rms=0.11 \\
B - g_{sm} &= -0.20 + 0.760~(g-r)_{sm},~rms=0.07 \\
I - i_{sm} &= ~~0.02 - 0.02~(r-i)_{sm},~rms=0.08,
\end{split}
\end{align*} 
the EQMOD system:
\begin{align*} 
\begin{split}
L - r_{sm} &= -0.13 + 0.38~(g-r)_{sm},~rms=0.18 \\
\end{split}
\end{align*} 
the iTelescope 33 system:
\begin{align*} 
\begin{split}
R - r_{sm} &= ~~0.07 - 0.17~(g-r)_{sm},~rms=0.04 \\
G - r_{sm} &= -0.29 + 0.79~(g-r)_{sm},~rms=0.05\\
B - g_{sm} &= -0.25 + 0.72~(g-r)_{sm},~rms=0.05,
\end{split}
\end{align*}
and finally, the CCESO system: 
\begin{align*} 
\begin{split}
R - r_{sm} &= -0.12 + 0.30~(g-r)_{sm},~rms=0.03\\
G - r_{sm} &= -0.16 + 0.40~(g-r)_{sm},~rms=0.02 \\
B - g_{sm} &= -0.17 - 0.53~(g-r)_{sm},~rms=0.03
\end{split}
\end{align*}

\section{Host galaxy and extinction}  
\label{sec:analysis_host}

SN~2024ggi occurrs in NGC~3621, a bulgeless late-type (Sd) spiral galaxy \citep[][]{2009ApJ...690.1031B} situated in the Hydra constellation. The location of the explosion site of SN~2024ggi is 110$\arcsec$ away from the galactic center \citep[][]{xiang2024red}. NGC~3621 is face-on with an extended disk \citep[][]{2004AJ....128...16K} and has an inclination angle of 25$^{\circ}$ \citep[][]{2018MNRAS.480.1973K}. NGC~3621 is known to host an active galactic nucleus \citep[][]{2007ApJ...663L...9S,2009ApJ...700.1759G} which is powered by a supermassive black hole of mass $\lesssim$\,$3\times10^{6}$\,M$_{\odot}$ \citep[][]{2009ApJ...690.1031B}. Additionally, NGC~3621 has a stellar mass of $8.1\times10^{9}$\,M$_{\odot}$ \citep[][]{2016MNRAS.457.2122G} and an absolute $B$-band magnitude of $-20.07\pm0.23$ and star formation rate of $830\pm640$\,M$_{\odot}$\,Myr$^{-1}$ in its star forming nucleus \citep{2018MNRAS.480.1973K}.

Various methods are available to measure the distances of NGC~3621, including Cepheids, the Tully-Fisher relation, and the tip of the red giant branch (TRGB). However, considering the potential contamination from non-RGB populations in the TRGB method and the intrinsic scatter associated with the Tully-Fisher relation, we ultimately adopted the distance derived from Cepheids. 
The distance to the host galaxy NGC~3621 is d = $6.64\pm0.3$\,Mpc, with a distance modulus of $\mu=29.11\pm0.06$\,mag \citep{freedman2001final}. 

We consider the Milky Way foreground extinction of $E(B-V)_{\rm MW} = 0.070$\,mag \citep[][]{Schlafly2011} in the direction of SN~2024ggi. The presence of Na\,I D2 and D1 lines in the high-resolution KAST spectrum indicates host galaxy extinction and \citet[][]{2024arXiv240419006J} estimate the corresponding host galaxy extinction of $E(B-V)_{\rm host} = 0.084\pm0.018$\,mag. Thus, we consider a total (Milky Way + host) extinction of $E(B-V)_{\rm total} = 0.154\pm0.018$\,mag in our analysis.

\bibliography{sample631}{}
\bibliographystyle{aasjournal}

\end{document}

%% file: Commands.tex
\newcommand{\I}{{\sc i}}
\newcommand{\II}{{\sc ii}}
\newcommand{\III}{{\sc iii}}
\newcommand{\IV}{{\sc iv}}
\newcommand{\V}{{\sc v}}


%% file: Packages.tex
\usepackage[utf8]{inputenc}
\usepackage{CJK}
\usepackage{multirow}
\usepackage{booktabs}
\usepackage{subfigure}
\usepackage{graphicx}
\usepackage{amsmath}

%% file: Tables/SLT_phot.tex

\begin{table*}
    \renewcommand*{\arraystretch}{1.1}
    \centering
    \caption{
        Photometry and spectroscopy observational log of SN~2024ggi. Magnitudes have not corrected for the expected foreground and host extinction. The errors for the optical photometry are quoted to $1 \sigma$, while upper limits are quoted to $3 \sigma$ significance. T0 is at MJD = 60410.90 (explosion epoch). The magnitudes of citizen science images marked with $^{*}$ using non-$griz$ filters are converted to $g$, $r$, and $i$ using the formulae provided in the appendix\,\ref{app:citizen_image_calibration} .
    }
    \begin{tabular}{cccccc}
        \toprule

        \multicolumn{6}{c}{Imaging}       \\

        \midrule
        
        $T_{\rm start} - T_0$       &MJD       &Telescope       &  Instrument        &Filter        &Apparent magnitude     \\
        (days)          &          &                &time (s)               &              &(AB mag)               \\
        \midrule
        

        $-5.84$      &60405.063       &ATLAS              & ACAM          &$o$        &$>19.80$            \\
        $+0.24$      &60411.141       &ATLAS              & ACAM          &$o$        &$18.9\pm0.102$            \\
        $+0.58$      &60411.476       &Lulin/SLT         & Andor SDK2           &$u$        &$17.1\pm0.158$            \\ 
        $+0.58$      &60411.479       &Lulin/SLT         & Andor SDK2           &$g$        &$16.4\pm0.04$            \\ 
        $+0.59$      &60411.485       &Lulin/SLT         & Andor SDK2           &$r$        &$16.4\pm0.074$            \\ 
        $+0.59$      &60411.489       &Lulin/SLT         & Andor SDK2           &$i$        &$16.2\pm0.189$            \\ 
        $+0.59$      &60411.493       &Lulin/SLT         & Andor SDK2           &$z$        &$16.5\pm0.095$            \\ 
        $+0.69$      &60411.589       &Swift              & UVOT           &$UVW1$        &$16.7\pm0.05$            \\
        $+0.69$      &60411.591       &Swift              & UVOT           &$u$        &$16.25\pm0.05$            \\
        $+0.69$      &60411.592       &Swift              & UVOT           &$b$        &$16.02\pm0.05$            \\
        $+0.69$      &60411.592       &Swift              & UVOT           &$UVW2$        &$17.0\pm0.05$            \\
        $+0.70$      &60411.596       &Swift              & UVOT           &$v$        &$16.0\pm0.07$            \\
        $+0.70$      &60411.597       &Swift              & UVOT           &$UVM2$        &$17.0\pm0.06$            \\
        $+0.70$      &60411.605       &EQMOD ASCOM HEQ5/6      & Atik Cameras          &$Lum$        &$15.43\pm0.26^{*}$    \\ 
        
$+0.71$      &60411.606&    TAM/RC12      & ASI174          &$Bessel~I$     &$16.08\pm0.12^{*}$ \\ 

$+1.48$      &60412.384        &    iTelescope 33      & Apogee USB/Net          &$Red$        &$13.32\pm0.02^{*}$            \\ 
$+1.49$      &60412.388        &    iTelescope 33      & Apogee USB/Net          &$Green$        &$13.41\pm0.01^{*}$            \\ 
$+1.49$      &60412.391        &    iTelescope 33      & Apogee USB/Net          &$Blue$        &$13.16\pm0.01^{*}$            \\ 

$+1.66$      &60412.565&    TAM/RC12      & ASI174          &$Lum$        &$12.90\pm0.01^{*}$            \\ 

$+2.69$      &60413.588&    CCESO/Planewave CDK17      & QHY163M          &$Green$        &$12.21\pm0.01^{*}$            \\ 
$+2.69$      &60413.588&    CCESO/Planewave CDK17      & QHY163M          &$Red$        &$12.12\pm0.01^{*}$            \\ 
$+2.69$      &60413.588&    CCESO/Planewave CDK17      & QHY163M          &$Blue$        &$11.92\pm0.01^{*}$            \\ 

        $+5.43$      &60416.325       &Pan-STARRS1         & Giga Pixel Camera           &$i$        &$12.070 \pm0.003$            \\
        $+5.43$      &60416.325       &Pan-STARRS1         & Giga Pixel Camera           &$r$        &$12.020 \pm0.005$            \\
        $+5.43$      &60416.326       &Pan-STARRS1         & Giga Pixel Camera           &$y$        &$12.237 \pm0.013$            \\   
        $+5.43$      &60416.326       &Pan-STARRS1         & Giga Pixel Camera           &$z$        &$12.195 \pm0.008$            \\
        $+5.43$      &60416.327       &Pan-STARRS1         & Giga Pixel Camera           &$w$        &$12.067 \pm0.003$            \\

$+9.67$      &60420.573&    TAM/RC12      & ASI174          &$Green$        &$11.99\pm0.03^{*}$            \\ 
$+9.68$      &60420.575&    TAM/RC12      & ASI174          &$Blue$        &$11.97\pm0.03^{*}$            \\

        \midrule

        \multicolumn{6}{c}{Spectroscopy}       \\

        \midrule

        $T_{\rm start} - T_0$       &MJD       &Telescope            &Instrument              &Exp. time                 & Wavelength Range                   \\
        (days)                      &          &                     &      &(sec)      & ($\AA$)      \\
        \midrule

        $+0.70$            & 60411.608      & Lijiang 2.4m                 & YFOSC     & 1800     & 3611-8929                              \\
        
        $+9.64$            & 60420.539      & LOT      & LISA     & 1800        & 3700-8436     \\
        $+17.73$           & 60428.633      & LOT      & LISA      & $300\times3$        &   3700-8436 \\        
        \bottomrule
    \end{tabular}
    \label{tab:lc_spec_log}
\end{table*}
